\documentclass[12pt,letterpaper]{extarticle}
\usepackage[
  letterpaper,
  left=1.25in,
  right=1.25in,
  top=1.25in,
  bottom=1.25in,
  includefoot,
  headheight=0pt,
  headsep=0pt,
  footskip=0.5in
]{geometry}
              
\usepackage{bm}

\usepackage[authoryear,round]{natbib}
\renewcommand{\citeyearpar}[1]{(\citeyear{#1})}

\usepackage[titletoc,title]{appendix}

\usepackage[T1]{fontenc}

\newfont{\elevenpt}{ptmr8t at 11pt}
\usepackage{graphicx,color}

\usepackage{hyperref}

\usepackage{amssymb}
\usepackage{amsfonts}
\usepackage{amsmath}

\usepackage{tikz}
\usepackage{pgfplots}
\pgfplotsset{compat=1.18}

\def\therefore{\hbox{.\raise1ex\hbox{.}.}}
\def\because{\leavevmode\raise1ex\hbox{.}.\raise1ex\hbox{.}}

\def\qed{\ \  \rule{0.5em}{2ex}}

\newcommand{\R}{\mathbb{R}}
\newcommand{\N}{\mathbb{N}}

\newcommand{\Q}{\mathbb{Q}}

\newcommand{\bw}{\boldsymbol{w}}

\newcommand{\bx}{\boldsymbol{x}}

\newcommand{\by}{\boldsymbol{y}}

\newcommand{\ba}{\boldsymbol{a}}
\newcommand{\bb}{\boldsymbol{b}}

\newcommand{\bone}{\boldsymbol{1}}
\newcommand{\zero}{\boldsymbol{0}}

\newcommand{\bz}{\boldsymbol{z}}

\title{Separable social welfare evaluation without completeness\footnote{We are grateful to Kristof Bosmans, Kohei Kamaga, Takeshi Murooka, Daisuke Nakajima, Kensei Nakamura, Tadashi Sekiguchi, Koji Shirai, Norio Takeoka, Shohei Yanagita, John Weymark, and participants of Mathematical Economics 2025 Workshop, Tokyo Social Choice Theory Workshop, and Economic Theory Workshop in Morioka 2026 for invaluable discussions and suggestions. Financial supports from Hitotsubashi University are gratefully acknowledged.}}
\author{Kaname Miyagishima\thanks{Hitotsubashi University. Email: \texttt{kaname1128@gmail.com}}
\and Eisuke Uchida\thanks{Hitotsubashi University. Email: \texttt{eisuke.m.uchida@gmail.com}}
\and Kenta Yoshida\thanks{Hitotsubashi University. Email: \texttt{kentayoshida.econ@gmail.com}}
}

\begin{document}
\maketitle

\begin{abstract}
    In this paper, we study incomplete social evaluation of utility vectors to capture situations in which there is disagreement in society about the degree of inequality aversion.
    We characterize multi-prioritarian criteria using efficiency, equity, impartiality, separability, and a condition called continuous divisible completion condition.
    We also provide uniqueness results and investigate comparative inequality aversion.
\end{abstract}

\section{Introduction}
The vast majority of studies on social welfare functions assume completeness of social criteria, requiring that any two social situations are comparable. 
This assumption is useful, as it makes the analysis tractable. 
However, the requirement of completeness in social welfare evaluations is often regarded as too demanding, and there are situations where admitting incompleteness is more appropriate.
As Sen (\citeyear{Sen2017}, pp. xxx--xxxi) argues:

\begin{quote}
The admission of partial orderings vastly expands the applicability of social choice theory.
This makes it possible, for example, to arrive at practical solutions despite some remaining disagreement, since the partial ranking of agreement may allow us to do many useful things.
There may well be little hope of complete agreement, for example, on what to do in taking care of the global environment (or, more particularly, in trying to prevent global warming), or on what must be done urgently to try to curb global pandemics, or remove medical neglect across the world.
And yet we can, with adequate public discussion and active advocacy, hope to get agreement on partial remedies that need not await a complete resolution of all our differences.
\end{quote}
Following this idea, we study incomplete social welfare criteria for ranking utility vectors in finite societies when there is disagreement about the degree of inequality aversion.
Our objective is to provide a social criterion that can be used under only partial agreement on inequality aversion.

Even within a single society, attitudes toward inequality and redistribution are heterogeneous, and it is difficult to reach full consensus on a single appropriate degree of inequality aversion.
Alesina and Giuliano (\citeyear{Alesina2011}) emphasize that heterogeneous interests, beliefs about fairness, and culturally shaped expectations generate persistent disagreement over redistribution.
In such environments, relying on a unique degree of inequality aversion to assess social situations is not compelling.
Even in this situation, social evaluation under partial agreement remains important, and our aim is to provide a criterion that enables such evaluation.

In this paper, we consider a system of social quasi-orderings, which is a family assigning one social quasi-ordering to each finite society.
We require four standard axioms in literature the social choice theory: Strong Pareto, the Pigou--Dalton transfer principle, Impartiality, and Separability.
Under a certain technical condition called {\it Continuous Divisible Completion}, we characterize an incomplete prioritarian social criterion, which we call the \textit{multi-prioritarian criterion}, represented in each society by a family of continuous, strictly increasing, and concave functions.
We call these functions \textit{evaluation functions}, each of which expresses a view of distributive evaluation that the social evaluator should take into consideration.
The multi-prioritarian criterion compares utility vectors by unanimous agreement among the evaluation functions in the set.
We also show that the generalized Lorenz criterion and an incomplete counterpart of the generalized Gini criterion, obtained by requiring unanimity across all admissible weight vectors, coincide with the multi-prioritarian criterion generated by the set of all increasing concave functions.
Moreover, we also establish affine uniqueness when finite representing families remain equivalent on every finite replicated population.

Our result contrasts with the well-known fact that, when a social criterion satisfying the above axioms is complete, the set of evaluation functions is a singleton (Maskin \citeyear{Maskin1978}; Blackorby et al. \citeyear{Blackorby2005}).
Under the multi-prioritarian criterion, the multiplicity of evaluation functions can be interpreted as reflecting disagreements about the degree of inequality aversion.
The greater the variety of evaluation functions with differing levels of concavity, the wider the range of disagreement. 
Our result also differs from Borie's (\citeyear{Borie2016}) characterization of additively separable quasi-orderings. 
As explained in Section 7, Borie relies on solvability and related measurement-theoretic richness conditions, which are restrictive in our social-evaluation setting. 
We instead use Continuous Divisible Completion to obtain our characterization. 
The variable-population framework becomes important for our uniqueness results across replicated populations.

The organization of the paper is as follows.
After laying down the model in Section 2, we introduce the axioms in Section 3.
In Section 4, we present the characterization of the multi-prioritarian criterion and its uniqueness results.
In Section 5, we examine the implications of conditions on the measurability and interpersonal comparability of utility.
In Section 6, we study comparative inequality aversion.
In Section 7, the related literature is discussed.
The proofs of the results in Section 4 are given in the Appendix.

\section{The model}
Let $\mathcal I$ be a countably infinite set of possible individuals and let
\[ \mathcal N=\{N\subseteq\mathcal I\mid 2\leq |N|<\infty\}. \]
For each society $N\in\mathcal N$, let $X_N=\R^N$ be the set of utility vectors,or allocations, for $N$.
A {\it social criterion} is a family $\boldsymbol{\succsim}=(\succsim_N)_{N\in\mathcal N}$ of quasi-orderings, where each $\succsim_N$ is reflexive and transitive but need not be complete.
Its asymmetric and symmetric parts are denoted by $\succ_N$ and $\sim_N$, respectively.

Given $K\subseteq N$, $\bx\in\R^K$, and $\bz\in\R^{N\setminus K}$, write $(\bx,\bz)$ for the combined allocation in $X_N$.
For societies $N,M\in\mathcal N$, a bijection $\pi:N\to M$, and $\bx\in X_N$, define $\bx^\pi\in X_M$ by $x^\pi_{\pi(i)}=x_i$ for every $i\in N$.

For society $N \in \mathcal N$ and $k\in\N$, choose pairwise disjoint societies $N^1,\ldots,N^k$ and bijections $\pi_\ell:N\to N^\ell$.
Write $kN=N^1\cup\cdots\cup N^k$ for the corresponding $k$-fold replicated population.
For $\bx\in X_N$, its $k$-fold replicated allocation is
\[ \bx^{[k]}=\underbrace{(\bx,\ldots,\bx)}_{k\text{ copies}}\in X_{kN}, \]
where the $\ell$th copy of $\bx$ is assigned to $N^\ell$ according to $\pi_\ell$.

For $t\in\R$, let $\delta_t$ denote the Dirac measure at $t$.
For every $N\in\mathcal N$ and $\bx\in X_N$, define the counting measure
\[ \mu_{\bx}=\sum_{i\in N}\delta_{x_i}. \]
Let
\[
 \mathcal E_0=\left\{
  \sum_{j=1}^m a_j\delta_{t_j}
  \,\middle|\,
  m\in\N,\ a_j\in\R,\ t_j\in\R,\ \sum_{j=1}^m a_j=0
 \right\}.
\]
Equip $\mathcal E_0$ with the weak topology $\sigma(\mathcal E_0,C(\R))$, where $C(\R)$ is the vector space of continuous real-valued functions on $\R$.
For each $N\in\mathcal N$, define
\[
 D_N=\{\mu_{\bx}-\mu_{\by}\mid \bx,\by\in X_N\},
 \ \ \ 
 P_N=\{\mu_{\bx}-\mu_{\by}\mid \bx\succsim_N\by\}.
\]
The positive rational conic hull of $P_N$ is
\[
 P_N^c=\left\{
  \sum_{\ell=1}^m q_\ell\nu_\ell
  \,\middle|\,
  m\in\N,\ q_\ell\in\Q_+,\ \nu_\ell\in P_N
 \right\}.
\]

\section{Axioms}
In this section, we introduce axioms for our social criteria that are standard in social choice and welfare economics.
These axioms are imposed separately on every finite society.

The first axiom is a separability condition.
\begin{description}
\item[Separability.]
For every $N\in\mathcal N$, every $K\subseteq N$, all $\bx,\by\in\R^K$, and all $\bz,\bw\in\R^{N\setminus K}$,
\[
 (\bx,\bz)\succsim_N(\by,\bz)
 \Longleftrightarrow
 (\bx,\bw)\succsim_N(\by,\bw).
\]
\end{description}
This axiom requires that, when the utilities of individuals outside $K$ are unchanged between two allocations, their common utility levels should not affect the comparison of the utility levels assigned to the individuals in $K$.

The next axiom requires social evaluation to be independent of individual identities.
\begin{description}
\item[Impartiality.]
For every $N\in\mathcal N$, every $\bx\in X_N$, and every permutation $\pi:N\to N$, $\bx\sim_N\bx^\pi$.
\end{description}
This axiom requires that permuting the utility levels among individuals should not change the social evaluation of an allocation.

The next axiom, Strong Pareto, is a standard efficiency requirement that Pareto improvements should be socially desirable.
\begin{description}
\item[Strong Pareto.]
For every $N\in\mathcal N$ and all $\bx,\by\in X_N$, if $x_i\geq y_i$ for every $i\in N$, with a strict inequality for at least one individual, then $\bx\succ_N\by$.
\end{description}

Lastly, we state a well-known equity axiom.
\begin{description}
\item[Pigou--Dalton.]
For every $N\in\mathcal N$, all $\bx,\by\in X_N$, all distinct $i,j\in N$, and every $t>0$, if
\[
 x_i=y_i+t\leq y_j-t=x_j
\]
and $x_k=y_k$ for every $k\notin\{i,j\}$, then $\bx\succsim_N\by$.
\end{description}
This axiom requires that reducing inequality in utilities between two individuals, while preserving their utility ranking and their total utility, should be socially accepted.

\section{Characterization of the multi-prioritarian criterion}
Let $\boldsymbol{\mathcal G}=(\mathcal G_N)_{N\in\mathcal N}$ be a system of nonempty families of continuous, strictly increasing, and concave functions $g:\R\to\R$.
The \textit{multi-prioritarian criterion} represented by $\boldsymbol{\mathcal G}$ is defined by
\begin{equation}\label{(1)}
 \bx\succsim_{\mathcal G_N}\by
 \Longleftrightarrow
 \sum_{i\in N}g(x_i)\geq\sum_{i\in N}g(y_i)
 \ \ \text{for every }g\in\mathcal G_N
\end{equation}
for every $N\in\mathcal N$ and all $\bx,\by\in X_N$.
A social criterion is multi-prioritarian if it admits such a representing system.

We use the following technical condition to characterize multi-prioritarian criteria.
\begin{description}
\item[Continuous Divisible Completion.]
For every $N\in\mathcal N$,
\[ \overline{P_N^c}\cap D_N=P_N, \]
where the closure is taken in $\sigma(\mathcal E_0,C(\R))$.
\end{description}
This condition means that, if one forms the positive rational conic hull of the accepted utility-distribution differences and then extracts the part realizable by integer population counts in society $N$, one recovers exactly the original set $P_N$.
In particular, it implies
\[ P_N^c\cap D_N=P_N. \]
The operations defining $P_N^c$ take place in the vector space $\mathcal E_0$.
Thus accepted differences may be divided by positive integers as well as added and multiplied by nonnegative integers, without interpreting the resulting coefficients as literal fractional individuals.

Under Separability and Impartiality, the condition can be rewritten as follows.
For every $N\in\mathcal N$, every net $(\eta_\alpha)$ in $P_N^c$, and all $\bx,\by\in X_N$,
\[
 \eta_\alpha\longrightarrow\mu_{\bx}-\mu_{\by}
 \ \ \text{in }\sigma(\mathcal E_0,C(\R))
\]
implies $\bx\succsim_N\by$.
By Lemma A below, Separability and Impartiality imply that the comparison of two allocations depends only on their counting-measure difference.
The equivalence then follows from the characterization of closure by convergent nets and from $P_N\subseteq P_N^c\subseteq\overline{P_N^c}$.
Convergence in this topology means that
\[
 \int f\,d\eta_\alpha
 \longrightarrow
 \sum_{i\in N}f(x_i)-\sum_{i\in N}f(y_i)
 \ \ \ \text{for every }f\in C(\R).
\]
Since the elements of $P_N^c$ need not themselves be realizable by allocation pairs, this net formulation is a topological interpretation of the conic completion under Separability and Impartiality, not an ordinary behavioral continuity axiom concerning small changes in feasible allocations.

Together with Separability and Impartiality, Continuous Divisible Completion implies the usual closed-graph continuity of $\succsim_N$.
More precisely, suppose that $(\bx^m)$ and $(\by^m)$ are sequences in $X_N$ satisfying $\bx^m\longrightarrow\bx$ and $\by^m\longrightarrow\by$
coordinatewise, and $\bx^m\succsim_N\by^m$ for all $m \in \N$.
For every $f\in C(\R)$,
\[ \int f\,d(\mu_{\bx^m}-\mu_{\by^m})=\sum_{i\in N}f(x_i^m)-\sum_{i\in N}f(y_i^m)\longrightarrow\sum_{i\in N}f(x_i)-\sum_{i\in N}f(y_i). \]
Thus $\mu_{\bx^m}-\mu_{\by^m}\longrightarrow\mu_{\bx}-\mu_{\by}$
weakly.
Since every difference in the sequence belongs to $P_N\subseteq P_N^c$, the qualified net formulation above implies $\bx\succsim_N\by$.
Hence the graph
\[
 \{(\bx,\by)\in X_N\times X_N\mid \bx\succsim_N\by\}
\]
is closed in the ordinary Euclidean topology.

Even under Separability and Impartiality, the converse does not hold in general.
Ordinary closed-graph continuity considers only nets or sequences of differences that are themselves generated by accepted pairs in $X_N$.
Continuous Divisible Completion also considers nets in $P_N^c$ whose elements may be fractional rational combinations of several accepted differences and need not themselves belong to $D_N$.
It is therefore a continuity condition on the divisible conic extension of the social criterion, and not merely a continuity condition on the original relation $\succsim_N$.

The following is our main characterization result.
\begin{description}
\item[Proposition 1.]
Assume that Continuous Divisible Completion holds.
Then the following conditions are equivalent:
\begin{enumerate}
\item[(i)] $\boldsymbol{\succsim}$ is a multi-prioritarian criterion represented by a system $\boldsymbol{\mathcal G}$;
\item[(ii)] $\boldsymbol{\succsim}$ satisfies Separability, Impartiality, Strong Pareto, and Pigou--Dalton.
\end{enumerate}
\end{description}

Proposition 1 permits the representing family to depend on the society.
Thus disagreement about inequality aversion may vary with population size or composition, while each fixed-population relation retains an additively separable unanimous representation.
Continuous Divisible Completion supplies the closed convex cone needed for dual separation and is stronger than ordinary closed-graph continuity.\\

\noindent
{\bf Remark 1.}
Continuous Divisible Completion cannot be omitted from Proposition 1.
Let
\[
 \phi(t)=-e^{-t},
 \ \ \ 
 \chi(t)=-e^{-2t},
\]
and, for every $N\in\mathcal N$, define
\[
 \bx\succsim_N^L\by
 \Longleftrightarrow
 \left(
  \sum_{i\in N}\phi(x_i),
  \sum_{i\in N}\chi(x_i)
 \right)
 \geq_{\mathrm{lex}}
 \left(
  \sum_{i\in N}\phi(y_i),
  \sum_{i\in N}\chi(y_i)
 \right),
\]
where $\geq_{\mathrm{lex}}$ is the lexicographic order on $\R^2$.
This criterion satisfies Separability and Impartiality since the two aggregate sums are additive and symmetric.
It satisfies Strong Pareto and Pigou--Dalton since $\phi$ is strictly increasing and strictly concave.

Fix a two-person society and let $\by=(0,0)$ and $\bx=(-\log(3/2),\log2)$.
Also, for every sufficiently large $m$, let $\bx^m=\bigl(-\log(3/2-1/m),\log2\bigr)$.
Then
\[
 \sum_{i \in N}\phi(x_i^m)=-2+\frac1m>-2=\sum_{i \in N}\phi(y_i),
\]
so $\bx^m\succsim_N^L\by$ for every sufficiently large $m$.
However,
\[
 \sum_{i \in N}\phi(x_i)=\sum_{i \in N}\phi(y_i)=-2,
 \ \ \ 
 \sum_{i \in N}\chi(x_i)=-\frac52<-2=\sum_{i \in N}\chi(y_i),
\]
and hence $\bx\not\succsim_N^L\by$, although $\bx^m\to\bx$.
Thus the graph of $\succsim_N^L$ is not closed and Continuous Divisible Completion fails.
Since every multi-prioritarian criterion has a closed graph, this criterion is not multi-prioritarian.
Therefore the remaining four axioms do not characterize multi-prioritarian criteria without Continuous Divisible Completion. \qed\\

\noindent
{\bf Remark 2.}
Here we consider two criteria: the generalized Lorenz criterion and the unanimous intersection of the family of generalized Gini criteria indexed by $\bw\in\mathcal W_n$.
We show that both coincide with the multi-prioritarian criterion generated by the set of all increasing concave functions.

Fix $N\in\mathcal N$, put $n=|N|$, and write
\[
 z_{(1)}\leq\cdots\leq z_{(n)},
 \ \ \ 
 S_k(\bz)=\sum_{r=1}^k z_{(r)}
 \ \ (k=1,\ldots,n).
\]
Let
\[
 \mathcal W_n=\left\{
  \bw\in\R_+^n
  \,\middle|\,
  w_1\geq\cdots\geq w_n,\ \  \sum_{r=1}^n w_r=1
 \right\}.
\]
For all $\bx,\by\in X_N$, the following conditions are equivalent:
\begin{enumerate}
\item for every $\bw\in\mathcal W_n$,
\[
 \sum_{r=1}^n w_rx_{(r)}\geq\sum_{r=1}^n w_ry_{(r)};
\]
\item $S_k(\bx)\geq S_k(\by)$ for every $k=1,\ldots,n$;
\item for every continuous increasing concave function $g:\R\to\R$,
\[
 \sum_{i\in N}g(x_i)\geq\sum_{i\in N}g(y_i).
\]
\end{enumerate}
The equivalence of the first two conditions follows from summation by parts:
\[
 \sum_{r=1}^n w_rx_{(r)}=w_nS_n(\bx)+\sum_{k=1}^{n-1}(w_k-w_{k+1})S_k(\bx).
\]
Conversely, fix $k\in\{1,\ldots,n\}$ and define $\bw^k\in\mathcal W_n$ by $w_r^k=1/k$ for $r\leq k$ and $w_r^k=0$ for $r>k$.
The inequality in the first condition for $\bw^k$ is equivalent to $S_k(\bx)\geq S_k(\by)$.
For this equivalence, the relevant distributional analogue is second-order inverse stochastic dominance, studied by Muliere and Scarsini (\citeyear{Muliere1989}, Theorem 2).
Their Theorem 2 characterizes this order through unanimous comparisons over a class of rank-weighted criteria.

The equivalence between the second and third conditions follows from the following argument.
For $c\in\R$, let $h_c(t)=\min\{t,c\}$ and $H_{\bz}(c)=\sum_{i\in N}h_c(z_i)$.
Writing $S_0(\bz)=0$, we have
\[
 H_{\bz}(c)=\min_{0\leq k\leq n}\{S_k(\bz)+(n-k)c\},
 \qquad
 S_k(\bz)=\sup_{c\in\R}\{H_{\bz}(c)-(n-k)c\}.
\]
Hence the second condition is equivalent to $H_{\bx}(c)\geq H_{\by}(c)$ for every $c\in\R$.
Since each $h_c$ is continuous, increasing, and concave, the third condition implies these inequalities and therefore the second condition.
Conversely, on any compact interval containing the coordinates of $\bx$ and $\by$, every continuous increasing concave $g$ is a uniform limit of increasing concave piecewise-linear functions.
Each such function is a constant plus a nonnegative linear combination of $t$ and functions $h_c$.
The second condition gives $\sum_{i\in N}x_i\geq\sum_{i\in N}y_i$ and, by the preceding identities, $H_{\bx}(c)\geq H_{\by}(c)$ for every $c$.
Thus the inequality holds for each piecewise-linear approximation and, by uniform convergence, for $g$.
To obtain a representing family satisfying the strict-increase requirement in \eqref{(1)}, note that for every continuous increasing concave $g$ and every $\varepsilon>0$, $g_\varepsilon(t)=g(t)+\varepsilon t$ is strictly increasing and concave.
Taking $\varepsilon\downarrow0$ shows that restricting the representing family to strictly increasing functions does not change the ordering.
Hence the unanimous intersection of the generally nonseparable rank-weight functions has the additively separable representation in \eqref{(1)}.
This equivalence is the finite-population counterpart of the equivalence between generalized Lorenz dominance and second-degree stochastic dominance established by Thistle (\citeyear{Thistle1989}, Proposition 4).
Thistle's Proposition 6 further relates this order to unanimous welfare comparisons. \qed \\

We next explore the uniqueness of the multi-prioritarian representation.
For this purpose, we need to focus on \textit{population-independent} systems.
A system $\boldsymbol{\mathcal G}=(\mathcal G_N)_{N\in\mathcal N}$ is population-independent if $\mathcal G_{N}=\mathcal G_M$ for all $N, M \in \mathcal N$.
This condition says that the sets of evaluation functions are common in all societies.
Although population independence is quite restrictive, we later show in Remark 3 that it is necessary for our uniqueness results.

We first consider the case in which every representing family is finite.
This case is practically meaningful because each evaluation function may be interpreted as a distinct normative position concerning the appropriate degree of inequality aversion.
For example, governments evaluating redistributive policies ordinarily formulate and compare only finitely many policy positions, owing to limitations of information, resources, and human deliberation.
It is therefore reasonable to begin with finitely many basic evaluation functions underlying those positions.

Let $\boldsymbol{\mathcal U}=(\mathcal U_N)_{N\in\mathcal N}$ and $\boldsymbol{\mathcal V}=(\mathcal V_N)_{N\in\mathcal N}$
be finite representing systems for two multi-prioritarian criteria.
For every $N\in\mathcal N$, write
\[
 \mathcal U_N=\{u_{N,1},\ldots,u_{N,r_N}\},
 \ \ \ 
 \mathcal V_N=\{v_{N,1},\ldots,v_{N,s_N}\},
\]
\[
 \mathbf u_N(t)=(u_{N,1}(t),\ldots,u_{N,r_N}(t)),
 \ \ \ 
 \mathbf v_N(t)=(v_{N,1}(t),\ldots,v_{N,s_N}(t)).
\]
We say that the two systems are \textit{equivalent} if, for all $N\in\mathcal N$ and all $\bx,\by\in X_N$,
\[
 \bx\succsim_{\mathcal U_N}\by
 \Longleftrightarrow
 \bx\succsim_{\mathcal V_N}\by.
\]
For every $N\in\mathcal N$, define
\[
 \Gamma_{\mathcal V_N}=\left\{
  \sum_{j=1}^m\bigl(\mathbf v_N(x_j)-\mathbf v_N(y_j)\bigr)
  \,\middle|\,
  m\in\N,\ x_j,y_j\in\R
 \right\}.
\]

\begin{description}
\item[Proposition 2.] Suppose that $\boldsymbol{\mathcal U}$ and $\boldsymbol{\mathcal V}$ are population-independent systems of finite families representing multi-prioritarian criteria.
Then the following conditions are equivalent:
\begin{enumerate}
\item[(i)] $\boldsymbol{\mathcal U}$ and $\boldsymbol{\mathcal V}$ are equivalent;
\item[(ii)] for every $N\in\mathcal N$, there exist a linear map $A_N:\R^{s_N}\to\R^{r_N}$ and a vector $\boldsymbol{\beta}_N\in\R^{r_N}$ such that
\begin{equation}\label{(2)} \mathbf u_N(t)=A_N\mathbf v_N(t)+\boldsymbol{\beta}_N \ \ \ (t\in\R), \end{equation}
and
\begin{equation}\label{(3)}
 \mathbf d\geq\zero
 \Longleftrightarrow
 A_N\mathbf d\geq\zero
 \ \ \ (\mathbf d\in\Gamma_{\mathcal V_N}).
\end{equation}
\end{enumerate}
\end{description}
This proposition shows that the representation of the multi-prioritarian criterion is unique if and only if the evaluation functions in $\mathcal{U}_N$ are related to those in $\mathcal{V}_N$ by an affine transformation that preserves the unanimous ranking of allocations induced by these evaluation functions.
This affine relationship essentially preserves the cardinal information about inequality aversion embodied in the evaluation functions.
In particular, the special case where both sets are singletons reduces to the standard affine uniqueness of additively separable representations shown by Debreu (\citeyear{Debreu1959}).

The assumption that the representing families are finite also plays a technical role.
Equivalence of population-independent systems allows any two finite lists of utility levels to be extended to allocations in a common replicated population and compared there.
The resulting additive map acts between finite-dimensional spaces and can therefore be extended to the linear map appearing in Proposition 2.

However, under certain conditions, we can obtain another uniqueness result when the set of evaluation functions may be infinite, as shown in Proposition 3 below.\footnote{In a different framework of incomplete preferences over lotteries, Dubra et al. \citeyearpar{Dubra2004} provide an alternative notion of uniqueness for expected multi-utility representations. They characterize uniqueness by the closure of the convex hull of the multi-utility representation on a domain of lotteries.
However, their proof relies on the duality between expected utility functions and probability measures, and thus cannot be directly applied in our setting, which lacks such a duality structure between evaluation functions and utility vectors.}
In contrast to Proposition 2, this result identifies the cone generated by the evaluation functions modulo constant functions.
It therefore shows that the two representations embody the same views on inequality aversion through nonnegative combinations, rather than merely being connected by a linear transformation that may contain negative coefficients.
For any family $\mathcal H$ of evaluation functions, write
\[
 \langle\mathcal H\rangle
 =
 \operatorname{cone}(\mathcal H)+\operatorname{span}\{\mathbf1\},
\]
where $\mathbf1$ denotes the constant function equal to one.

\begin{description}
\item[Proposition 3.]
Suppose that  $\boldsymbol{\mathcal U}$ and $\boldsymbol{\mathcal V}$ are population-independent systems representing multi-prioritarian criteria.
Suppose there exist population-independent systems of finite families
\[
 \boldsymbol{\mathbb U}=(\mathbb U_N)_{N\in\mathcal N},
 \ \ \ 
 \boldsymbol{\mathbb V}=(\mathbb V_N)_{N\in\mathcal N}.
\]
For every $N\in\mathcal N$, let
\[
 \mathbb U_N=\{u_{N,1},\ldots,u_{N,r_N}\}\subseteq\mathcal U_N,
 \ \ \ 
 \mathbb V_N=\{v_{N,1},\ldots,v_{N,s_N}\}\subseteq\mathcal V_N,
\]
and suppose that
\[
 \langle\mathcal U_N\rangle=\operatorname{cone}(\mathbb U_N)+\operatorname{span}\{\mathbf 1\},
 \ \ \ 
 \langle\mathcal V_N\rangle=\operatorname{cone}(\mathbb V_N)+\operatorname{span}\{\mathbf 1\}.
\]
Then the following conditions are equivalent:
\begin{enumerate}
\item[(i)] $\boldsymbol{\mathcal U}$ and $\boldsymbol{\mathcal V}$ are equivalent.
\item[(ii)] $\langle\mathcal U_N\rangle=\langle\mathcal V_N\rangle$ for every $N\in\mathcal N$.
\end{enumerate}
\end{description}

Proposition 3 states that two possibly infinite representing systems generate the same social criterion if and only if they generate the same cone modulo constant functions in every society, provided that these cones admit finite population-independent generating systems.
The members of $\mathbb U_N$ and $\mathbb V_N$ may be interpreted as basic measurements of inequality aversion from which all evaluations relevant to the unanimous criterion are generated.
As in Proposition 2, restricting the basic measurements to a finite number is both practically meaningful and technically useful, even though the represented families $\mathcal U_N$ and $\mathcal V_N$ may themselves be infinite.\\

\noindent
{\bf Remark 3.}
Here, we show that population independence cannot in general be omitted from Propositions 2 and 3.
For $a>0$, define $u_a(t)=-e^{-at}$.
Fix a two-person society $N^0$ and define $\mathcal U_{N^0}=\{u_1,u_3\}$ and $V_{N^0}=\{u_1,u_2,u_3\}$.
For every $M\in\mathcal N\setminus\{N^0\}$, let $\mathcal U_M=\mathcal V_M=\{u_1,u_2,u_3\}$.
These are systems of finite families of continuous, strictly increasing, and strictly concave functions, but $\boldsymbol{\mathcal U}$ is not population-independent.

We first show that the two systems are equivalent.
For a positive two-dimensional vector $\bz=(z_1,z_2)$, write
$S_a(\bz)=z_1^a+z_2^a$.
The identities
\[
 S_3(\bz)=S_1(\bz)^3-3z_1z_2S_1(\bz)
 \quad\text{and}\quad
 S_2(\bz)=S_1(\bz)^2-2z_1z_2
\]
give
\[
 S_2(\bz)
 =\frac13S_1(\bz)^2
 +\frac23\frac{S_3(\bz)}{S_1(\bz)}.
\]
Consequently, for positive two-dimensional vectors $\bz$ and $\bw$,
\[
 S_1(\bz)\leq S_1(\bw)
 \quad\text{and}\quad
 S_3(\bz)\leq S_3(\bw)
 \quad\Longrightarrow\quad
 S_2(\bz)\leq S_2(\bw).
\]
Indeed, if $F(s,q)=s^2/3+2q/(3s)$, then $F(s,q)$ is increasing in $q$, and it is increasing in $s$ on $[s_0,\infty)$ whenever $q\leq s_0^3$.
Applying this observation first with $q=S_3(\bz)$ and then with $s=S_1(\bw)$ proves the implication.
For allocations $\bx,\by\in X_{N^0}$, apply the implication to $z_i=e^{-x_i}$ and $w_i=e^{-y_i}$.
It follows that the aggregate inequality for $u_2$ is implied by those for $u_1$ and $u_3$.
Hence $\succsim_{\mathcal U_{N^0}}=\succsim_{\mathcal V_{N^0}}$.
Since the two families coincide in every other society, $\boldsymbol{\mathcal U}$ and $\boldsymbol{\mathcal V}$ are equivalent.

Nevertheless, the conclusion of Proposition 2 fails at $N^0$.
Let
\[
 \bz=\left(\frac12,\frac52,\frac52\right),
 \qquad
 \bw=\left(1,\frac32,3\right).
\]
Direct calculation gives
\[
 S_1(\bz)=S_1(\bw),
 \qquad
 S_3(\bz)=S_3(\bw),
 \qquad
 S_2(\bz)-S_2(\bw)=\frac12.
\]
Put $x_j=-\log z_j$ and $y_j=-\log w_j$ for $j=1,2,3$, and order the coordinates of $\mathbf v_{N^0}$ as $(u_1,u_2,u_3)$.
Then
\[
 \mathbf d
 =\sum_{j=1}^3\bigl(\mathbf v_{N^0}(x_j)-\mathbf v_{N^0}(y_j)\bigr)
 =\left(0,-\frac12,0\right)
 \in\Gamma_{\mathcal V_{N^0}}.
\]
If a linear map $A_{N^0}$ and a vector $\boldsymbol\beta_{N^0}$ satisfied \eqref{(2)}, summing \eqref{(2)} over the two lists and cancelling the affine constants would give
\[
 A_{N^0}\mathbf d
 =\sum_{j=1}^3\bigl(\mathbf u_{N^0}(x_j)-\mathbf u_{N^0}(y_j)\bigr)
 =(0,0).
\]
Thus $A_{N^0}\mathbf d\geq\zero$, whereas $\mathbf d\not\geq\zero$, contradicting \eqref{(3)}.
Therefore equivalence does not imply condition (ii) of Proposition 2 without population independence.

The conclusion of Proposition 3 also fails.
Taking the finite generating families to be the representing families themselves, equivalence holds, but
\[
 \operatorname{cone}(\mathcal U_{N^0})+\operatorname{span}\{\mathbf1\}
 \neq
 \operatorname{cone}(\mathcal V_{N^0})+\operatorname{span}\{\mathbf1\}.
\]
Indeed, $u_2$ does not belong to the left-hand side.
Otherwise, there would exist $a,b\geq0$ and $c\in\R$ such that $-e^{-2t}=-ae^{-t}-be^{-3t}+c$ for all $t \in \R$.
Setting $z=e^{-t}$ would give the polynomial identity $-z^2=-az-bz^3+c$ for all $z>0$,
which is impossible.
Hence equivalence does not imply equality of the generated cones modulo constant functions without population independence.

\section{Measurability and interpersonal comparability of utility}
\label{sec:comparability}

In this section, we investigate the implications of measurability and
interpersonal comparability of utility for the multi-prioritarian criterion.
For complete prioritarian orderings, familiar comparability conditions
restrict the form of the evaluation function.
We examine how these conclusions extend when social comparisons require
agreement among a family of evaluation functions.

Each comparability condition below is imposed on each quasi-ordering 
$\succsim_N$ and does not introduce comparisons between allocations in
different societies $N$ and $N'$.
For every $N\in\mathcal N$, let $\mathbf 1_N\in\R^N$ denote the vector whose
components are all equal to one.
Throughout, we retain the multi-prioritarian representation characterized
in Proposition 1.
The proofs are given in Appendix B.

\subsection{Translation-scale Full Comparability}

The first condition means that the origin of the utility scale has no
normative significance and can be translated arbitrarily.
\begin{description}
\item[Population-wise Translation-scale Full Comparability (P-TFC).]
For each $N\in\mathcal N$ and $b\in\R$, the following holds for all
$\bx,\by\in X_N$:
\[
 \bx\succsim_N\by
 \Longleftrightarrow
 \bx+b\mathbf 1_N
 \succsim_N
 \by+b\mathbf 1_N.
\]
\end{description}
Under this condition, there exists a common,
interpersonally comparable unit with which
we can measure people's well-being, and the numerical difference
between the utility levels of any two individuals is meaningful.

It is known in the literature that, for complete prioritarian orderings with fixed population size,
Translation-scale Full Comparability
leads to the Kolm--Pollak criterion (Blackorby and Donaldson
\citeyear{Blackorby1982}; Blackorby et al. \citeyear{Blackorby2005})
defined by the following function $\kappa_\theta$ with some parameter $\theta\in\R_+$:
\begin{description}
\item[Definition.]
For every $\theta\in\R_+$, define the Kolm--Pollak evaluation function
$\kappa_\theta:\R\to\R$ by
\[
 \kappa_\theta(t)
 =
 \begin{cases}
 -e^{-\theta t} & \text{if }\theta>0,\\
 t & \text{if }\theta=0.
 \end{cases}
\]
The corresponding Kolm--Pollak criterion for society $N$ compares
allocations $\bx\in X_N$ by the value of $\sum_{i\in N}\kappa_\theta(x_i)$
\end{description}
For a multi-prioritarian criterion, P-TFC is imposed on the social
criterion $\succsim_N$ represented by agreement among evaluation functions.
The following proposition shows that, under the finite-generation
condition of Proposition 3, a social criterion satisfying P-TFC admits 
a representation by a finite family of Kolm--Pollak evaluation functions.
The corresponding population-independent parameter system is unique.

\begin{description}
\item[Proposition 4.]
Let $\boldsymbol{\mathcal G}=(\mathcal G_N)_{N\in\mathcal N}$ be a
population-independent system representing $\boldsymbol{\succsim}$.
Suppose that it admits a population-independent system of nonempty finite
generating families in  the sense of Proposition 3.
Then the following conditions are equivalent:
\begin{enumerate}
\item[(i)] $\boldsymbol{\succsim}$ satisfies P-TFC 
\item[(ii)] there
exists a population-independent system
$\boldsymbol\Theta=(\Theta_N)_{N\in\mathcal N}$ of nonempty finite
sets $\Theta_N\subseteq\R_+$ such that, for every $N\in\mathcal N$
and all $\bx,\by\in X_N$,
\[
 \bx\succsim_N\by
 \Longleftrightarrow
 \sum_{i\in N}\kappa_\theta(x_i)
 \geq
 \sum_{i\in N}\kappa_\theta(y_i)
 \quad\text{for every }\theta\in\Theta_N.
\]
\end{enumerate}
Moreover, this parameter system $\boldsymbol\Theta$ is unique.
\end{description}

When a social criterion admits such a representation, it can therefore be
understood as requiring agreement among a finite collection of Kolm--Pollak
criteria, each associated with a degree of inequality aversion.
The set $\Theta_N$ identifies these degrees of ineqality aversion.

Note that the finite-generation condition cannot be omitted. 
P-TFC alone does not guarantee a representation by a finite set of Kolm--Pollak evaluation functions.
Indeed, the generalized Lorenz criterion in Remark 2 satisfies P-TFC. 
However, for any society with at least two individuals, it admits no representation by a finite family of continuous, strictly increasing, concave evaluators.\footnote{
    Suppose that a finite family $\{g_1,\ldots,g_r\}$ 
    of continuous, strictly increasing, concave evaluation functions
    represents the generalized Lorenz criterion. 
    Choose a point $t$ at which all $g_j$ are differentiable, 
    since such a point exists by concavity. 
    Strict monotonicity and concavity imply $g_j'(t)>0$. 
    For sufficiently small $\varepsilon>0$,
    \[
        g_j(t-\varepsilon)+g_j(t+\varepsilon)-2g_j(t-\varepsilon/2)
        =\varepsilon g_j'(t)+o(\varepsilon)>0
        \text{ for all } j.
    \]
    Thus the finite family ranks
    $(t-\varepsilon,t+\varepsilon)$ above
    $(t-\varepsilon/2,t-\varepsilon/2)$, whereas the generalized
    Lorenz criterion does not. 
    This contradicts the assumption when $|N|=2$.
    For a society with more than two individuals, assign utility $t$ to every remaining individual in both allocations.
    These common terms cancel in every additive comparison, while the first allocation still has a strictly lower minimum utility.
    Then we obtain the desired contradiction.
}

\subsection{Ratio-scale Full Comparability}

The next condition means that we can arbitrarily change the unit or scale of
the well-being measure without altering our moral judgment.
For this subsection, we restrict the domain to $X_N=\R_{++}^N$ for every
$N\in\mathcal N$; the preceding definitions are understood on this
restricted domain.
\begin{description}
\item[Population-wise Ratio-scale Full Comparability (P-RFC).]
For every $N\in\mathcal N$, all $\bx,\by\in X_N$, and every $a>0$,
\[
 \bx\succsim_N\by
 \Longleftrightarrow
 a\bx\succsim_N a\by.
\]
\end{description}
This condition presumes a normatively significant fixed origin of well-being that is common among all individuals.
As argued by Blackorby and Donaldson (\citeyear{Blackorby1982}, p. 253), ``Ratio-scale Full Comparability involves picking an interpersonally significant norm such as a poverty line or a utility level at which life is no more worthwhile than death.''
For complete prioritarian orderings, Ratio-scale Full Comparability leads to
the generalized mean form (Blackorby and Donaldson
\citeyear{Blackorby1982}; Blackorby et al. \citeyear{Blackorby2005}) 
defined by the following function $\nu_\gamma$ with some parameter $\gamma\in(-\infty,1]$:
\begin{description}
\item[Definition.]
For every $\gamma\in(-\infty,1]$, define an evaluation
function $\nu_\gamma:\R_{++}\to\R$ by
\[
 \nu_\gamma(t)
 =
 \begin{cases}
 t^\gamma & \text{if }0<\gamma\leq1,\\
 \log t & \text{if }\gamma=0,\\
 -t^\gamma & \text{if }\gamma<0.
 \end{cases}
\]
The corresponding generalized mean criterion of order $\gamma$ for society
$N$ compares allocations $\bx\in X_N$ by $\sum_{i\in N}\nu_\gamma(x_i)$
\end{description}

As in the previous subsection, the following proposition shows that, under the finite-generation condition of Proposition 3 adapted to $\R_{++}$, 
a criterion satisfying P-RFC admits a representation by a finite family of
generalized mean evaluation functions. 
The corresponding parameter system is unique.

\begin{description}
\item[Proposition 5.]
Suppose that $X_N=\R_{++}^N$ for every $N\in\mathcal N$.
Let $\boldsymbol{\mathcal G}=(\mathcal G_N)_{N\in\mathcal N}$ be a
population-independent system representing $\boldsymbol{\succsim}$.
Suppose that it admits a population-independent system of nonempty finite
generating families as in the analogue of Proposition 3 on $\R_{++}$
Then the following conditions are equivalent:
\begin{enumerate}
\item[(i)] $\boldsymbol{\succsim}$ satisfies P-RFC 
\item[(ii)] there
exists a population-independent system
$\boldsymbol\Gamma=(\Gamma_N)_{N\in\mathcal N}$ of nonempty finite
sets $\Gamma_N\subseteq(-\infty,1]$ such that, for every $N\in\mathcal N$
and all $\bx,\by\in X_N$,
\[
 \bx\succsim_N\by
 \Longleftrightarrow
 \sum_{i\in N}\nu_\gamma(x_i)
 \geq
 \sum_{i\in N}\nu_\gamma(y_i)
 \quad\text{for every }\gamma\in\Gamma_N.
\]
\end{enumerate}
Moreover, this parameter system $\boldsymbol\Gamma$ is unique.
\end{description}

When a P-RFC criterion admits such a representation, it can therefore be
understood as requiring agreement among a finite collection of generalized
mean criteria, each associated with a degree of inequality aversion.  The set
$\Gamma_N$ identifies these degrees.

Again, the finite-generation condition is important. 
P-RFC alone does not guarantee a representation by a finite set of generalized mean evaluation functions, 
because the generalized Lorenz criterion restricted
to $\R_{++}^N$, satisfies P-RFC.

\subsection{Cardinal Full Comparability}

We now return to $X_N=\R^N$ for every $N\in\mathcal N$. 
Translation-scale Full Comparability and Ratio-scale Full Comparability may
be restrictive because utility does not necessarily have a common unit or
origin among individuals.
The next condition means that we can arbitrarily change both the origin and
the unit of the well-being measure. 
If only cardinally comparable utilities are available, the following condition may be imposed.

\begin{description}
\item[Population-wise Cardinal Full Comparability (P-CFC).]
For every $N\in\mathcal N$, every $a>0$, and every $b\in\R$, the following
holds for all $\bx,\by\in X_N$:
\[
 \bx\succsim_N\by
 \Longleftrightarrow
 a\bx+b\mathbf 1_N
 \succsim_N
 a\by+b\mathbf 1_N.
\]
\end{description}

For complete prioritarian orderings, Cardinal Full Comparability leads to the
utilitarian criterion (Maskin \citeyear{Maskin1978}). 
As in the preceding two cases, under the finite-generation
condition of Proposition 3, P-CFC yields the utilitarian criterion.

\begin{description}
\item[Proposition 6.]
Suppose that $X_N=\R^N$ for every $N\in\mathcal N$.
Let $\boldsymbol{\mathcal G}=(\mathcal G_N)_{N\in\mathcal N}$ be a
population-independent system representing $\boldsymbol{\succsim}$.
Suppose that it admits a population-independent system of nonempty finite
generating families in the sense of Proposition 3.
Then the following conditions are equivalent:
\begin{enumerate}
\item[(i)] $\boldsymbol{\succsim}$ satisfies P-CFC 
\item[(ii)] for every $N\in\mathcal N$ and all $\bx,\by\in X_N$,
\[
 \bx\succsim_N\by
 \Longleftrightarrow
 \sum_{i\in N}x_i\geq\sum_{i\in N}y_i.
\]
\end{enumerate}
\end{description}

Proposition 6 shows that, under population independence and finite
generation, P-CFC forces the multi-prioritarian criterion to be
utilitarian, and hence complete.
The generalized Lorenz criterion illustrates why finite generation condition is needed. 
It satisfies P-CFC. Nevertheless, it is generally incomplete and differs from
utilitarian.\footnote{
    It is known that cardinal full comparability alone does not lead to utilitarian. 
    The intersection of the utilitarian and leximin orderings studied by Blackorby and Donaldson 
    (\citeyear{BlackorbyDonaldson1977}) provides an earlier example. 
    See also Kamaga (\citeyear{Kamaga2018}). 
    For a fixed population, this relation is 
    invariant under common positive affine transformations, since both constituent orderings are invariant. However, it is generally incomplete and not utilitarian.
}
For example, $(0,10)$ and $(2,2)$ are incomparable under the
generalized Lorenz criterion, whereas the utilitarian criterion
strictly prefers the former.

\section{Comparative Inequality Aversion}
In this section, for simplicity, we fix an arbitrary society $N\in\mathcal N$ and write $n=|N|$ and $X=X_N$,
suppressing the population subscript when no confusion can arise.
In this section, we consider comparative inequality aversion of multi-prioritarian criteria.
Given two multi-prioritarian criteria $\succsim$ represented by $\mathcal G$ and $\succsim'$ represented by $\mathcal G'$, we say that $\succsim$ is more inequality averse than $\succsim'$ if the following (a) and (b) hold.
\begin{align*}
   \text{(a)}\ &\forall t\in \R
   ,\ \forall\bx \in X, \ \bx \succsim t\bone \Longrightarrow  \bx \succsim' t\bone ;\\
   \text{(b)}\ &\forall t\in \R
   ,\ \forall\bx \in X, \ t\bone \succ' \bx \Longrightarrow t\bone \succ \bx.
\end{align*}
 Condition (a) mentions that if the more inequality averse criterion $\succsim$ accepts the inequality in $\by$ compared to the equal situation $x\bone$, the less inequality averse criterion $\succsim'$ also accepts the inequality.
 Condition (b) states that if $\succsim'$ prefers an equal allocation to an unequal allocation, $\succsim$ also prefers the former to the latter.
Since our social criteria are quasi-orderings, we adopt these two conditions.
If we just consider (a), even if $\succsim'$ weakly prefers an equal situation $x \bone$ to an unequal one $\by$, those situations may be incomparable for $\succsim$.
Similarly, if we only introduce (b), even when $\succsim$ accepts an strictly unequal allocation over an equal one, $\succsim'$ may not be able to compare these two allocations.

Let $\mathcal G$ be a nonempty family representing $\succsim$.
For each evaluation function $g\in\mathcal G$, define
\[
c_g(\by)=g^{-1}\left(\frac1n\sum_{i\in N}g(y_i)\right).
\]
This is the \emph{equally distributed equivalent} utility level associated with $g$: 
the allocations $\bx$ and $c_g(\bx)\bone$ have the same additive evaluation under $g$.
Since each $g\in\mathcal G$ is continuous and strictly increasing,
$c_g(\bx)$ is well defined and $\min_{i\in N}x_i\leq c_g(\bx)\leq\max_{i\in N}x_i$.

These equally distributed equivalent utility levels determine comparisons with equal allocations:
\[
\bx\succsim t\bone
\Longleftrightarrow
\frac1n\sum_{i\in N}g(x_i)\ge g(t) \text{ for every } g\in\mathcal G,
\Longleftrightarrow
\inf_{g\in\mathcal G}c_g(\bx)\ge t.
\]
Since $t\bone\succ\bx$ is equivalent to 
$t\bone\succsim\bx$ and $\bx\not\succsim t\bone$, we have
\[
t\bone\succ\bx
\Longleftrightarrow
t \ge \sup_{g\in\mathcal G}c_g(\bx) \text{ and } t > \inf_{g\in\mathcal G}c_g(\bx).
\]
Consequently,
\[
\inf_{g\in\mathcal G}c_g(\bx)=\max\{t\in\R:\bx\succsim t\bone\},
\quad
\sup_{g\in\mathcal G}c_g(\bx)=\min\{t\in\R:t\bone\succsim\bx\}.
\]
These right hand sides depend only on $\succsim$, even when its representing family is not unique.

The next result characterizes comparative inequality aversion in terms of equally distributed equivalent utility levels.
\begin{description}
    \item[Proposition 7.] Let $\succsim$ and $\succsim'$ be multi-prioritarian criteria represented by $\mathcal G$ and $\mathcal G'$, respectively. The following statements are equivalent.
    \begin{description}
        \item[(i)] $\succsim$ is more inequality averse than $\succsim'$.
        \item[(ii)] For every $\bx\in X$, $\displaystyle\inf_{g'\in\mathcal G'}c_{g'}(\bx)\geq\inf_{g\in\mathcal G}c_g(\bx)$
        and $\displaystyle\sup_{g'\in\mathcal G'}c_{g'}(\bx)\geq\sup_{g\in\mathcal G}c_g(\bx)$.
    \end{description}
\end{description}

{\it Proof.} 
First, suppose (i) holds and fix $\bx\in X$.
Condition (a) implies
$\inf_{g'\in\mathcal G'}c_{g'}(\bx)\geq\inf_{g\in\mathcal G}c_g(\bx)$.
If $\sup_{g'\in\mathcal G'}c_{g'}(\bx)<\sup_{g\in\mathcal G}c_g(\bx)$, 
choose $t$ between these two values.
Then $\sup_{g \in\mathcal G}c_g(\bx) > t > \sup_{g'\in\mathcal G'}c_{g'}(\bx) \geq \inf_{g\in\mathcal G}c_g(\bx)$ holds 
and this implies $t\bone\succ'\bx$ but $t\bone\not\succ\bx$, contradicting condition (b).
Thus (ii) holds.

Conversely, suppose (ii) holds.
If $\bx\succsim t\bone$, then
$t\leq\inf_{g\in\mathcal G}c_g(\bx)\leq\inf_{g'\in\mathcal G'}c_{g'}(\bx)$, so $\bx\succsim' t\bone$.
If $t\bone\succ'\bx$, then $t\geq\sup_{g'\in\mathcal G'}c_{g'}(\bx)\geq\sup_{g\in\mathcal G}c_g(\bx)$ and 
$t>\inf_{g'\in\mathcal G'}c_{g'}(\bx)\geq\inf_{g\in\mathcal G}c_g(\bx)$ hold and this implies $t\bone\succ\bx$.
This proves (i). \qed

Proposition 7 provides a simple criterion for comparing the inequality aversion of two multi-prioritarian criteria.
That is, the result shows that one multi-prioritarian criterion is more inequality averse than another if and only if, for every allocation, both the infimum and the supremum of its equally distributed equivalent utility levels are weakly lower under the former criterion.

The following corollary provides a sufficient condition 
for comparative inequality aversion in terms of 
concavity of evaluation functions.
\footnote{
    The comparison of evaluation functions used here is adapted from 
    \citeauthor{CerreiaVioglioOk2018} (\citeyear{CerreiaVioglioOk2018}, Proposition 5.6), 
    who study comparative risk aversion for possibly incomplete preferences over lotteries.
}
\begin{description}
    \item[Corollary 8.] Let $\succsim$ and $\succsim'$ be multi-prioritarian criteria represented by $\mathcal G$ and $\mathcal G'$, respectively. Suppose that
    \begin{description}
        \item[(i)] for every $g'\in\mathcal G'$, there exists $g\in\mathcal G$ such that $g$ is more concave than $g'$
        \footnote{
            We say that $g$ is \emph{more concave than} $g'$ if there exists a nondecreasing concave function $\phi:\R\to\R$ such that $g=\phi\circ g'$.
        } and
        \item[(ii)] for every $g\in\mathcal G$, there exists $g'\in\mathcal G'$ such that $g$ is more concave than $g'$.
    \end{description}
    Then $\succsim$ is more inequality averse than $\succsim'$.
\end{description}
{\it Proof.} Suppose $g\in\mathcal G$ and $g'\in\mathcal G'$ satisfy $g=\phi\circ g'$ for a nondecreasing concave function $\phi:\R\to\R$. Jensen's inequality gives
\[
g\left(c_g(\bx)\right)
=\frac1n\sum_{i\in N}\phi\!\left(g'(x_i)\right)
\leq\phi\!\left(\frac1n\sum_{i\in N}g'(x_i)\right)
=\phi\!\left(g'\left(c_{g'}(\bx)\right)\right)
=g\!\left(c_{g'}(\bx)\right).
\]
Since $g$ is strictly increasing, this implies $c_g(\bx)\leq c_{g'}(\bx)$ for every $\bx\in X$.
By (i), we have $\inf_{g\in\mathcal G}c_g(\bx)\leq\inf_{g'\in\mathcal G'}c_{g'}(\bx)$.
By (ii), we have $\sup_{g\in\mathcal G}c_g(\bx)\leq\sup_{g'\in\mathcal G'}c_{g'}(\bx)$.
Now, Proposition 7 implies the conclusion. \qed

\section{Related Literature}

We discuss the related literature.
This paper builds on two strands of literature, namely, the prioritarian social welfare ranking over utility distributions in social choice theory and incomplete preferences in decision theory.
 
In social choice theory, Maskin \citeyearpar{Maskin1978} characterizes a class of generalized utilitarian criteria by applying Debreu's \citeyearpar{Debreu1959} additive separability representation theorem. 
A generalized utilitarian criterion with a symmetric and concave evaluation function is called "prioritarian" because it prioritizes the benefit of the worse-off individual. 
The philosophical idea of prioritarianism was proposed by Parfit \citeyearpar{Parfit1991} in his Lindley Lecture. 
The class of prioritarian criteria is widely used in welfare economics and normative theory (see, among others, Blackorby et al. \citeyearpar{Blackorby2005} and Adler and Norheim \citeyearpar{AdlerNorheim2022}).
 Chung and Miyagishima \citeyearpar{ChungMiyagishima2024} formulated Parfit's (\citeyear{Parfit1991}) priority view as an axiom for social welfare orderings and examined its implications. \par

There are some papers that study incomplete social welfare criteria.
For instance, several studies by Cato (\citeyear{Cato2018}, \citeyear{Cato2020}) present versions of Arrow's impossibility results under incompleteness of social preference.
Danan et al. \citeyearpar{Danan2015} considered a preference aggregation problem under risk in the case of incomplete preferences at the individual and social level.
Danan et al. \citeyearpar{Danan2016} and Nakamura \citeyearpar{Nakamura2025} studied preference aggregation under uncertainty when agents and society have incomplete preferences.
Pivato (\citeyear{Pivato2013}) considered incomplete social evaluation when utilities are ordinally interpersonally comparable, while Pivato (\citeyear{Pivato2015}) studied incomplete criteria to aggregate individual welfare gains.
To the best of our knowledge, however, this paper is the first to examine incomplete social criteria to rank utility vectors when there may be disagreements on the degree of inequality aversion.

The mathematical foundation of our result is related to the theory of multi-utility representations of quasi-orderings.
The pioneering work by Aumann \citeyearpar{Aumann1962} relaxed the completeness axiom to pursue a von Neumann--Morgenstern-style representation, and Evren and Ok \citeyearpar{Evren2011} comprehensively studied multi-utility representations of incomplete preferences.
Our proof instead constructs, for each society, a closed cone of accepted utility-distribution differences and separates that cone by continuous linear functionals.\par
Within the standard model of choice over lotteries, the representation of incomplete preferences has been explored through various models focusing on the indeterminacy of vNM functions, often referred to as multi-taste models.
For instance, Dubra et al. \citeyearpar{Dubra2004} demonstrated that incomplete preferences over lotteries on a compact metrizable space can be represented by expected utility with a set of vNM functions.
Our framework is different from these models since we do not have a structure of duality between evaluation functions in $\mathcal G$ and allocations in $X$.

Formally, the closest paper to ours is Borie's \citeyearpar{Borie2016}. Under different settings and conditions, his Theorems 2 and 3 provide additively separable representations of quasi-orderings, which are quite similar to multi-prioritarian criteria.
Among the conditions in Borie's Theorems 2 and 3 is Restricted Solvability.
To state this condition in our setting, fix $N\in\mathcal{N}$ and consider $\succsim_N$ on $X_N$.
For $\bx\in X_N$, $i\in N$, and $v_i\in\R$, let $\bx_{-i}v_i$ denote the allocation obtained from $\bx$ by replacing $x_i$ with $v_i$.
\begin{description}
\item[Restricted Solvability.] If $\bx,\by\in X_N$ and $u_i,w_i\in \R$ satisfy $\bx_{-i}u_i\succsim_N\by\succsim_N\bx_{-i}w_i$, then there exists $v_i\in \R$ such that $\bx_{-i}v_i\sim_N\by$.
\end{description}
We argue that this condition is too restrictive for social evaluations that accommodate multiple views of inequality aversion, because the condition rules out criteria represented by evaluation functions that are not positive affine transformations of each other.
To see this, let $g_1,g_2\in\mathcal G_N$ be continuous and strictly increasing, and suppose that they are not positive affine transformations of each other.
Then there exist $s,t\in \R$, with $s<t$, such that, for
\[
c=g_1^{-1}\left(\frac{g_1(s)+g_1(t)}{2}\right),
\]
we have $2g_2(c)\neq g_2(s)+g_2(t)$.
Otherwise, writing $\varphi=g_2\circ g_1^{-1}$, we would have
\[
\varphi\left(\frac{u+v}{2}\right)=\frac{\varphi(u)+\varphi(v)}{2}
\quad\text{for all }u,v\in g_1(\R).
\]
Then, continuity implies that $\varphi$ is affine.
Since $\varphi$ is strictly increasing, this would make $g_2$ a positive affine transformation of $g_1$, contrary to the assumption.

Suppose that our criterion satisfies Restricted Solvability.
Choose $\bx,\by\in X_N$ that agree in all coordinates other than $i$ and $j$, with $x_j=c$, $y_i=s$, and $y_j=t$.
Since $s\leq c\leq t$, we have $\bx_{-i}t_i\succsim_N\by\succsim_N\bx_{-i}s_i$.
Restricted Solvability would yield $v_i\in \R$ such that $\bx_{-i}v_i\sim_N\by$.
Indifference under $g_1$ requires $g_1(v_i)+g_1(c)=g_1(s)+g_1(t)$, so $v_i=c$.
Indifference under $g_2$ would then require $2g_2(c)=g_2(s)+g_2(t)$, a contradiction.
Hence the criterion does not satisfy Restricted Solvability.\footnote{A similar problem arises with another condition in Borie (\citeyear{Borie2016}), but we omit the argument for brevity.}

The above arguments show that Borie's (\citeyear{Borie2016}) results are not suitable for our purposes since Restricted Solvability is incompatible with representations by sets of evaluation functions with different degrees of inequality aversion.
In contrast, our characterization results can avoid this problem and are therefore more useful for social evaluation that accommodates diverse views on inequality aversion.

\section{Appendix A: Proofs of Results in Section 4}

We first record a fixed-population consequence of Separability and Impartiality.

\begin{description}
\item[Lemma A.]
For every $N\in\mathcal N$ and all $\bx,\by,\bx',\by'\in X_N$, if $\mu_{\bx}-\mu_{\by}=\mu_{\bx'}-\mu_{\by'}$,
then
\[
 \bx\succsim_N\by\Longleftrightarrow\bx'\succsim_N\by'.
\]
\end{description}

\noindent
{\it Proof.}
Define positive counting measures
\[ \lambda=\sum_{t\in\R}\min\{\mu_{\bx}(\{t\}),\mu_{\by}(\{t\})\}\delta_t 
 \ \ \text{ and } \ \  
 \lambda'=\sum_{t\in\R}\min\{\mu_{\bx'}(\{t\}),\mu_{\by'}(\{t\})\}\delta_t.
\]
Then it is straightforward to show
\[ \mu_{\bx}(\{t\})-\lambda(\{t\})=\max\{\mu_{\bx}(\{t\})-\mu_{\by}(\{t\}),0\}. \]
and
\[ \mu_{\by}(\{t\})-\lambda(\{t\})=\max\{\mu_{\by}(\{t\})-\mu_{\bx}(\{t\}),0\}. \]
Similar equations obviously hold for $\bx'$, $\by'$, and $\lambda'$.
Since $\mu_{\bx}-\mu_{\by}=\mu_{\bx'}-\mu_{\by'}$, for every $t\in\R$ we have $\mu_{\bx}(\{t\})-\mu_{\by}(\{t\})=\mu_{\bx'}(\{t\})-\mu_{\by'}(\{t\})$.
Thus, we obtain
\[
 \mu_{\bx}-\lambda=\mu_{\bx'}-\lambda'
 \ \ \text{and}\ \ 
 \mu_{\by}-\lambda=\mu_{\by'}-\lambda'.
\]
Put $r=(\mu_{\bx}-\lambda)(\R)=(\mu_{\by}-\lambda)(\R)$. 
Then $r\in\N\cup\{0\}$ and $\lambda(\R)=\lambda'(\R)=|N|-r$.
Choose $K\subseteq N$ with $|K|=r$, allocations $\ba,\bb\in\R^K$ with counting measures $\mu_{\bx}-\lambda$ and $\mu_{\by}-\lambda$, and allocations $\bz,\bw\in\R^{N\setminus K}$ with counting measures $\lambda$ and $\lambda'$.
Impartiality permits the four original allocations to be replaced by $(\ba,\bz)$, $(\bb,\bz)$, $(\ba,\bw)$, and $(\bb,\bw)$, respectively.
By Separability, we obtain the desired conclusion. \qed \\ 

\noindent
{\it Proof of Proposition 1.}
It is straightforward to check that (i) implies (ii).
We prove that (ii) implies (i). 
Suppose $\succsim$ satisfies the axioms in (ii).
Fix $N\in\mathcal N$.
By Lemma A, for all $\bx,\by\in X_N$,
\begin{equation}\label{(4)}
 \bx\succsim_N\by
 \Longleftrightarrow
 \mu_{\bx}-\mu_{\by}\in P_N.
\end{equation}
Let $C_N=\overline{P_N^c}\subseteq\mathcal E_0$.
This is a weakly closed convex cone containing zero.
By Continuous Divisible Completion,
\begin{equation}\label{(5)} C_N\cap D_N=P_N. \end{equation}
Combining \eqref{(4)} and \eqref{(5)},
\begin{equation}\label{(6)}
 \bx\succsim_N\by
 \Longleftrightarrow
 \mu_{\bx}-\mu_{\by}\in C_N.
\end{equation}

For $g\in C(\R)$ and $\nu=\sum_{j=1}^m a_j\delta_{t_j}\in\mathcal E_0$, define
\[
 \int g\,d\nu=\sum_{j=1}^m a_jg(t_j).
\]
Let
\[ C_N^*=\left\{g\in C(\R)\,\middle|\,\int g\,d\nu\geq0 \text{ for every }\nu\in C_N\right\}. \]
Since every $\nu\in\mathcal E_0$ has total mass zero, every constant function $t\mapsto c$ satisfies
\[
 \int c\,d\nu=c\nu(\R)=0
 \ \ (\nu\in\mathcal E_0).
\]
Let $\mathbf 1:\R\to\R$ denote the constant function defined by $\mathbf 1(t)=1$ for every $t\in\R$, and let $\R\mathbf 1=\{c\mathbf 1\mid c\in\R\}$.
The quotient space $C(\R)/\R\mathbf 1$ consists of the equivalence classes under the relation
\[
 g\sim h
 \Longleftrightarrow
 g-h\in\R\mathbf 1.
\]
For $g\in C(\R)$, its equivalence class is
\[
 [g]=g+\R\mathbf 1
 =\{g+c\mathbf 1\mid c\in\R\}.
\]
Such functions are identified because $\int(g+c\mathbf 1)\,d\nu=\int g\,d\nu$ for every $\nu\in\mathcal E_0$.
Consider the dual pair
\[ \left\langle\mathcal E_0,C(\R)/\R\mathbf 1\right\rangle, \ \ \langle\nu,[g]\rangle=\int g\,d\nu. \]
This pairing separates both spaces: continuous functions separate finitely supported signed measures, while a nonconstant $g$ is separated from the constant class by some $\delta_s-\delta_t$.
The induced weak topology on $\mathcal E_0$ is $\sigma(\mathcal E_0,C(\R))$.

The inclusion of $C_N$ in the right-hand side of the following equality follows directly from the definition of $C_N^*$.
For the reverse inclusion, fix $\nu\notin C_N$.
Since $C_N$ is convex and weakly closed, by the separation theorem for locally convex spaces (Aliprantis and Border \citeyearpar{AliprantisBorder2006}, Corollary~5.80), there exist a weakly continuous linear functional $L$ on $\mathcal E_0$ and $\alpha\in\R$ such that
\[
 L(\nu)<\alpha\leq L(\eta)
 \ \ \text{for every }\eta\in C_N.
\]
Since $0\in C_N$, one has $\alpha\leq0$.
Suppose, to the contrary, that $L(\eta)<0$ for some $\eta\in C_N$.
Because $C_N$ is a cone, $r\eta\in C_N$ for every $r>0$, whereas $L(r\eta)<\alpha$ for sufficiently large $r$, a contradiction.
Hence $L(\eta)\geq0$ for every $\eta\in C_N$.
Combining $L(\nu)<\alpha$ with $\alpha\leq0$, we obtain $L(\nu)<0$.
The continuous dual of $\bigl(\mathcal E_0,\sigma(\mathcal E_0,C(\R)/\R\mathbf 1)\bigr)$ is $C(\R)/\R\mathbf 1$.
Hence there exists $g\in C(\R)$ such that
\[
 L(\eta)=\int g\,d\eta
 \ \ \text{for every }\eta\in\mathcal E_0.
\]
It follows that $g\in C_N^*$ and $\int g\,d\nu<0$.
Therefore,
\begin{equation}\label{(7)} C_N=\left\{\nu\in\mathcal E_0\,\middle|\,\int g\,d\nu\geq0 \text{ for every }g\in C_N^*\right\}. \end{equation}

Fix $a,b\in\R$ with $a<b$.
By Strong Pareto and \eqref{(6)}, we have $\delta_b-\delta_a\in C_N$.
Hence every $g\in C_N^*$ is nondecreasing.
By Pigou--Dalton and \eqref{(6)}, we have $2\delta_{(a+b)/2}-\delta_a-\delta_b\in C_N$.
Hence every $g\in C_N^*$ is midpoint concave (that is, $g((x+y)/2)\geq(g(x)+g(y))/2$ for every $x,y\in\R$).
By continuity, every member of $C_N^*$ is concave.

It remains to obtain strict increasingness.
For every rational pair $p<q$, Strong Pareto and \eqref{(6)} imply
$\delta_q-\delta_p\in C_N$ and $\delta_p-\delta_q\notin C_N$.
By \eqref{(7)}, there exists $g_{p,q}\in C_N^*$ such that $g_{p,q}(q)>g_{p,q}(p)$.
Enumerate the rational pairs as $(p_m,q_m)_{m\geq1}$, normalize $g_m=g_{p_m,q_m}-g_{p_m,q_m}(0)$, and define\footnote{A related construction of a strictly increasing function by a countable weighted sum appears in the proof of Proposition~5.2(i) of McCarthy et al. \citeyearpar{McCarthy2021}.
They rescale their functions to ensure that $|L_i(v_j)|\leq1$ whenever $j\leq i$, and then sum them with weights $2^{-i}$.
The normalization here is different: countability comes from the rational subintervals of $\R$, and $M_m$ controls $g_m$ on the expanding compact intervals $[-m,m]$.
This is needed because $\R$ is noncompact and ensures uniform convergence on compact intervals.}
\[
 M_m=\max_{|t|\leq m}|g_m(t)|,
 \ \ \ 
 \lambda_m=\frac{2^{-m}}{1+M_m},
 \ \ \ 
 h_N(t)=\sum_{m=1}^{\infty}\lambda_mg_m(t).
\]
The series converges uniformly on compact intervals.
Thus $h_N$ is continuous.
Let
\[
 h_{N,k}=\sum_{m=1}^k\lambda_mg_m.
\]
Since $C_N^*$ is a convex cone, $\lambda_m>0$, and $g_m\in C_N^*$, one has $h_{N,k}\in C_N^*$ for every $k$.
Fix $\nu\in C_N$.
Its support is finite and is contained in $[-r,r]$ for some $r>0$.
Uniform convergence on $[-r,r]$ implies
\[ \int h_N\,d\nu=\lim_{k\to\infty}\int h_{N,k}\,d\nu. \]
Each integral on the right is nonnegative because $h_{N,k}\in C_N^*$ and $\nu\in C_N$.
Hence $\int h_N\,d\nu\geq0$.
Since this holds for every $\nu\in C_N$, it follows that $h_N\in C_N^*$.
It is concave and nondecreasing.
If $x<y$, choose rationals $x<p_m<q_m<y$.
Then
\[
 h_N(y)-h_N(x)
 \geq
 \lambda_m\bigl(g_m(q_m)-g_m(p_m)\bigr)
 >0.
\]
Thus $h_N$ is strictly increasing.

Define $\mathcal G_N=\{g+\varepsilon h_N\mid g\in C_N^*,\ \varepsilon>0\}$.
Every member of $\mathcal G_N$ is continuous, strictly increasing, and concave.
If $\nu\notin C_N$, by \eqref{(7)}, there exists $g\in C_N^*$ with $\int g\,d\nu<0$.
Choosing $\varepsilon>0$ sufficiently small preserves this inequality for $g+\varepsilon h_N$.
Hence
\[ C_N=\left\{\nu\in\mathcal E_0\,\middle|\,\int g\,d\nu\geq0 \text{ for every }g\in\mathcal G_N\right\}. \]
Combining this equality with \eqref{(6)} proves $\succsim_N=\succsim_{\mathcal G_N}$.
Since $N$ was arbitrary, condition (i) follows. \qed\\

\noindent
{\it Proof of Proposition 2.}
Suppose condition (i) holds.
Fix $N\in\mathcal N$.
Define
\[ L_{\mathcal V_N}=\operatorname{span}\{\mathbf v_N(x)-\mathbf v_N(y)\mid x,y\in\R\}. \]
Fix $m\in\N$ and two lists $(x_1,\ldots,x_m),(y_1,\ldots,y_m)\in\R^m$ such that
\[
 \sum_{j=1}^m\mathbf v_N(x_j)=\sum_{j=1}^m\mathbf v_N(y_j).
\]
Choose $k$ with $k|N|\geq m$, fix $t_0\in\R$, and put $p=k|N|-m$.
Extend both lists to length $k|N|$ as follows:
\[
 (x_1,\ldots,x_m,\underbrace{t_0,\ldots,t_0}_{p\text{ copies}})
 \ \ \text{and}\ \ 
 (y_1,\ldots,y_m,\underbrace{t_0,\ldots,t_0}_{p\text{ copies}}).
\]
Since the systems $\boldsymbol{\mathcal V}$ and $\boldsymbol{\mathcal U}$ are population-independent, $\mathcal V_{kN}=\mathcal V_N$ and $\mathcal U_{kN}=\mathcal U_N$.
These lists are considered as allocations in $X_{kN}$ that are indifferent with respect to $\succsim_{\mathcal V_{kN}}$.
By condition (i), $\succsim_{\mathcal V_{kN}}=\succsim_{\mathcal U_{kN}}$.
Hence they are also indifferent with respect to $\succsim_{\mathcal U_{kN}}$.
By the definition of $\succsim_{\mathcal U_{kN}}$,
\[
 \sum_{j=1}^m\mathbf u_N(x_j)+p\mathbf u_N(t_0)=\sum_{j=1}^m\mathbf u_N(y_j)+p\mathbf u_N(t_0).
\]
Subtracting $p\mathbf u_N(t_0)$ from both sides gives
\begin{equation}\label{(8)}
 \sum_{j=1}^m\mathbf v_N(x_j)=\sum_{j=1}^m\mathbf v_N(y_j)
 \ \ \Longrightarrow\ \ 
 \sum_{j=1}^m\mathbf u_N(x_j)=\sum_{j=1}^m\mathbf u_N(y_j).
\end{equation}

For
\[
 \mathbf d=\sum_{j=1}^m\bigl(\mathbf v_N(x_j)-\mathbf v_N(y_j)\bigr)\in\Gamma_{\mathcal V_N},
\]
define
\[
 A_N(\mathbf d)=\sum_{j=1}^m\bigl(\mathbf u_N(x_j)-\mathbf u_N(y_j)\bigr).
\]
By \eqref{(8)}, this definition is independent of the representation of $\mathbf d$.
Then for
\[
 \mathbf d=\sum_{j=1}^m\bigl(\mathbf v_N(x_j)-\mathbf v_N(y_j)\bigr)
 \ \ \text{and}\ \ 
 \mathbf d'=\sum_{\ell=1}^{m'}\bigl(\mathbf v_N(x'_{\ell})-\mathbf v_N(y'_{\ell})\bigr),
\]
we have
\[
 A_N(\mathbf d+\mathbf d')
 =\sum_{j=1}^m\bigl(\mathbf u_N(x_j)-\mathbf u_N(y_j)\bigr)
 +\sum_{\ell=1}^{m'}\bigl(\mathbf u_N(x'_{\ell})-\mathbf u_N(y'_{\ell})\bigr)
 =A_N(\mathbf d)+A_N(\mathbf d').
\]
Thus $A_N$ is additive.

We show that $A_N:\Gamma_{\mathcal V_N}\to\R^{r_N}$ is continuous at zero, where $\Gamma_{\mathcal V_N}$ has the subspace topology inherited from $L_{\mathcal V_N}$.
Fix $a,b\in\R$ with $a<b$ and put $\mathbf e=\mathbf v_N(b)-\mathbf v_N(a)\gg\zero$.
There is a neighborhood $W$ of zero in $L_{\mathcal V_N}$ such that $\mathbf e+\mathbf d\geq\zero$ and $\mathbf e-\mathbf d\geq\zero$ for every $\mathbf d\in W$.
Fix $\mathbf d\in\Gamma_{\mathcal V_N}\cap W$.
For each $\eta\in\{1,-1\}$, there exist $m_\eta\in\N$ and two lists $(x_1^\eta,\ldots,x_{m_\eta}^\eta),(y_1^\eta,\ldots,y_{m_\eta}^\eta)\in\R^{m_\eta}$ such that
\[
 \mathbf e+\eta\mathbf d=\sum_{j=1}^{m_\eta}\bigl(\mathbf v_N(x_j^\eta)-\mathbf v_N(y_j^\eta)\bigr).
\]
Choose $k_\eta\in\N$ with $k_\eta|N|\geq m_\eta$ and extend both lists by $k_\eta|N|-m_\eta$ copies of $t_0$.
The extended lists are allocations in $X_{k_\eta N}$.
Since $\mathbf e+\eta\mathbf d\geq\zero$, the first allocation is weakly preferred to the second with respect to $\succsim_{\mathcal V_{k_\eta N}}$.
Since $\boldsymbol{\mathcal U}$ and $\boldsymbol{\mathcal V}$ are population-independent and condition (i) implies $\succsim_{\mathcal V_{k_\eta N}}=\succsim_{\mathcal U_{k_\eta N}}$, the first allocation is also weakly preferred to the second with respect to $\succsim_{\mathcal U_{k_\eta N}}$.
By the definition of $A_N$,
\[
 A_N(\mathbf e+\mathbf d)\geq\zero,
 \ \ \ 
 A_N(\mathbf e-\mathbf d)\geq\zero,
\]
and hence
\[
 -A_N(\mathbf e)\leq A_N(\mathbf d)\leq A_N(\mathbf e).
\]
Fix $\varepsilon>0$.
Put $M=1+\|A_N(\mathbf e)\|_\infty$ and choose $q\in\N$ with $q>M/\varepsilon$.
If $\mathbf z\in\Gamma_{\mathcal V_N}\cap q^{-1}W$, then $q\mathbf z\in\Gamma_{\mathcal V_N}\cap W$.
Therefore,
\[
 -A_N(\mathbf e)\leq A_N(q\mathbf z)\leq A_N(\mathbf e).
\]
Since $A_N$ is additive, $A_N(q\mathbf z)=qA_N(\mathbf z)$.
Hence
\[
 \|A_N(\mathbf z)\|_\infty
 \leq\frac{\|A_N(\mathbf e)\|_\infty}{q}
 <\varepsilon.
\]
Thus $A_N:\Gamma_{\mathcal V_N}\to\R^{r_N}$ is continuous at zero.

We next show that $\Gamma_{\mathcal V_N}$ is dense in $L_{\mathcal V_N}$.
Indeed, the set $\{\mathbf v_N(x)-\mathbf v_N(y)\mid x,y\in\R\}$ is connected, contains zero, and spans $L_{\mathcal V_N}$; the connected component of zero in the closure of the generated additive group is a vector subspace and therefore equals $L_{\mathcal V_N}$.
Hence $A_N$ has a unique continuous additive extension to $L_{\mathcal V_N}$, which is real linear.
Extend it linearly to $\R^{s_N}$.
The extension need not be unique outside $L_{\mathcal V_N}$.

For every $t\in\R$,
\[ \mathbf u_N(t)-\mathbf u_N(t_0)=A_N\bigl(\mathbf v_N(t)-\mathbf v_N(t_0)\bigr). \]
Setting $\boldsymbol{\beta}_N=\mathbf u_N(t_0)-A_N\mathbf v_N(t_0)$ proves \eqref{(2)}.
Fix $\mathbf d\in\Gamma_{\mathcal V_N}$.
Choose $m\in\N$ and two lists $(x_1,\ldots,x_m),(y_1,\ldots,y_m)\in\R^m$ such that
\[
 \mathbf d=\sum_{j=1}^m\bigl(\mathbf v_N(x_j)-\mathbf v_N(y_j)\bigr).
\]
Choose $k\in\N$ with $k|N|\geq m$ and extend both lists by $k|N|-m$ copies of $t_0$.
Using $\mathcal V_{kN}=\mathcal V_N$, $\mathcal U_{kN}=\mathcal U_N$, and $\succsim_{\mathcal V_{kN}}=\succsim_{\mathcal U_{kN}}$, the definitions of the two criteria imply
\[
 \mathbf d\geq\zero\Longleftrightarrow A_N\mathbf d\geq\zero,
\]
which is \eqref{(3)}.
Summing \eqref{(2)} over $X_N$, we have the aggregate formula.
Since $N$ was arbitrary, condition (ii) follows.

Conversely, suppose condition (ii) holds.
Fix $N\in\mathcal N$ and let $\bx,\by\in X_N$.
Set
\[ \mathbf d=\sum_{i\in N}\mathbf v_N(x_i)-\sum_{i\in N}\mathbf v_N(y_i)\in\Gamma_{\mathcal V_N}. \]
The affine constants cancel.
By \eqref{(3)},
\[
 \bx\succsim_{\mathcal V_N}\by
 \Longleftrightarrow
 \mathbf d\geq\zero
 \Longleftrightarrow
 A_N\mathbf d\geq\zero
 \Longleftrightarrow
 \bx\succsim_{\mathcal U_N}\by.
\]
Since $N$ was arbitrary, condition (i) holds. \qed

\medskip
\noindent
{\it Proof of Proposition 3.}
Suppose condition (ii) holds.
Fix $N\in\mathcal N$.
Nonnegative combinations preserve every aggregate inequality, and constant functions cancel between allocations of equal population size.
Hence the criteria generated by $\mathcal U_N$ and $\mathcal V_N$ agree on $X_N$.
Since $N$ was arbitrary, condition (i) follows.

Conversely, suppose condition (i) holds.
Finite generation implies that the population-independent systems $\boldsymbol{\mathbb U}$ and $\boldsymbol{\mathbb V}$ are equivalent.
We can therefore apply Proposition 2 to $\boldsymbol{\mathbb U}$ and $\boldsymbol{\mathbb V}$.
Fix $N\in\mathcal N$ and write
\[
 \mathbf u_N(t)=(u_{N,1}(t),\ldots,u_{N,r_N}(t)),
 \ \ \ 
 \mathbf v_N(t)=(v_{N,1}(t),\ldots,v_{N,s_N}(t)).
\]
There exist matrices $A_N\in\R^{r_N\times s_N}$ and $B_N\in\R^{s_N\times r_N}$ and vectors $\boldsymbol{\alpha}_N\in\R^{r_N}$ and $\boldsymbol{\beta}_N\in\R^{s_N}$ such that
\[
 \mathbf u_N(t)=A_N\mathbf v_N(t)+\boldsymbol{\alpha}_N,
 \ \ \ 
 \mathbf v_N(t)=B_N\mathbf u_N(t)+\boldsymbol{\beta}_N.
\]
Let $\mathbf a_{N,j}$ be the $j$th row of $A_N$, and define
\[ L_{\mathbb V_N}=\operatorname{span}\{\mathbf v_N(x)-\mathbf v_N(y)\mid x,y\in\R\}. \]
The order-preservation conclusion of Proposition 2 implies
\[
 \mathbf a_{N,j}\mathbf d\geq0
 \ \ \ (\mathbf d\in\Gamma_{\mathbb V_N}\cap\R_+^{s_N}).
\]
We claim that $\Gamma_{\mathbb V_N}\cap\R_+^{s_N}$ is dense in $L_{\mathbb V_N}\cap\R_+^{s_N}$.
To see this, fix $\mathbf d\in L_{\mathbb V_N}\cap\R_+^{s_N}$ and $t_0\in\R$, and let $y_q=t_0+1/q$.
Since every function in $\mathbb V_N$ is continuous and strictly increasing,
\[ \mathbf h_q=\mathbf v_N(y_q)-\mathbf v_N(t_0)\gg\zero \ \ \text{and}\ \ \mathbf h_q\to\zero \ \ \text{ as }q\to\infty. \]
By the density of $\Gamma_{\mathbb V_N}$ in $L_{\mathbb V_N}$ (shown in the proof of Proposition 2), we have $\mathbf d_q\in\Gamma_{\mathbb V_N}$ such that
\[
 \|\mathbf d_q-\mathbf d\|_\infty<\min_m(\mathbf h_q)_m.
\]
where $(\mathbf h_q)_m$ denotes the $m$th coordinate of $\mathbf h_q$.
Then $\mathbf d_q+\mathbf h_q\in\Gamma_{\mathbb V_N}\cap\R_+^{s_N}$ and $\mathbf d_q+\mathbf h_q\to\mathbf d$.
Hence $\Gamma_{\mathbb V_N}\cap\R_+^{s_N}$ is dense in $L_{\mathbb V_N}\cap\R_+^{s_N}$.
By continuity of the linear functional $\mathbf d\mapsto\mathbf a_{N,j}\mathbf d$, we have $\mathbf a_{N,j}\mathbf d\geq0$ for every $\mathbf d\in L_{\mathbb V_N}\cap\R_+^{s_N}$.
Since $\mathbf h_q\in L_{\mathbb V_N}\cap\R_{++}^{s_N}$, one has $L_{\mathbb V_N}\cap\operatorname{int}\R_+^{s_N}\neq\varnothing$.
By Corollary 16.4.2 of Rockafellar (\citeyear{Rockafellar1970}), we obtain
\[ \bigl( L_{\mathbb V_N}\cap\R_+^{s_N} \bigr)^*=L_{\mathbb V_N}^{\perp}+\R_+^{s_N}. \]
where $\bigl( L_{\mathbb V_N}\cap\R_+^{s_N} \bigr)^*$
denotes the dual cone of $L_{\mathbb V_N}\cap\R_+^{s_N} $.
Hence $\mathbf a_{N,j}=\mathbf b_{N,j}+\mathbf c_{N,j}$ for some $\mathbf b_{N,j}\in\R_+^{s_N}$ and $\mathbf c_{N,j}\in L_{\mathbb V_N}^{\perp}$.
Since $\mathbf c_{N,j}\in L_{\mathbb V_N}^{\perp}$, one has $\mathbf c_{N,j}\mathbf v_N(t)=\mathbf c_{N,j}\mathbf v_N(t_0)$ for every $t\in\R$.
Thus
\[ u_{N,j}(t)=\mathbf b_{N,j}\mathbf v_N(t)+\bigl(\mathbf c_{N,j}\mathbf v_N(t_0)+\alpha_{N,j}\bigr). \]
Since $\mathbf b_{N,j}\in\R_+^{s_N}$, we have $ u_{N,j}\in\operatorname{cone}(\mathbb V_N)+\operatorname{span}\{\mathbf 1\}$ 
for every $j=1,\ldots,r_N$.
The same argument applied to $B_N$ shows that each $v_{N,m}$ belongs to $\operatorname{cone}(\mathbb U_N)+\operatorname{span}\{\mathbf 1\}$.
Therefore,
\[ \operatorname{cone}(\mathbb U_N)+\operatorname{span}\{\mathbf 1\}=\operatorname{cone}(\mathbb V_N)+\operatorname{span}\{\mathbf 1\}. \]
Hence $\langle\mathcal U_N\rangle=\langle\mathcal V_N\rangle$.
Since $N$ was arbitrary, condition (ii) follows. \qed

\section{Appendix B: Proofs of Results in Section 5}
\label{app:comparability-proofs}

\noindent
{\it Proof of Proposition 4.}
The implication (ii)$\Rightarrow$(i) is immediate,
since $\kappa_\theta(\cdot+b)$ is a positive affine
transformation of $\kappa_\theta$ for every
$\theta\geq0$ and $b\in\R$.

Suppose that (i) holds.
Choose a population-independent system
$\boldsymbol{\mathbb G}=(\mathbb G_N)_{N\in\mathcal N}$
of nonempty finite generating families in the sense of Proposition 3.
This system also represents $\boldsymbol{\succsim}$. 
Thus, for every $N\in\mathcal N$ and all $\bx,\by\in X_N$,
\[
 \bx\succsim_N\by
 \Longleftrightarrow
 \bx\succsim_{\mathbb G_N}\by.
\]
Let $[g]=g+\R\mathbf1\in C(\R)/\R\mathbf1$ denote the equivalence class of
$g$ modulo constant functions.
Define $\mathcal K_N=\operatorname{cone}\{[g]\mid g\in\mathbb G_N\}$.
Delete generators whose classes are nonnegative combinations of the
remaining classes until no further deletion is possible.
This procedure terminates because
the original family is finite and does not change $\mathcal K_N$.
The remaining classes represent the distinct extreme rays of \(\mathcal K_N\), 
since nontrivial nonnegative combinations of strictly increasing functions cannot be constant.

Because $\mathbb G_{N}=\mathbb G_M$ and hence
$\mathcal K_{N}=\mathcal K_M$ for all $N,M\in\mathcal{N}$, the same deletion and labeling can be used
for all populations, so the resulting system remains population
independent.  Henceforth, write
$\mathbb G_N=\{g_{N,1},\ldots,g_{N,r_N}\}$ for the resulting family,
whose classes represent the distinct extreme rays of $\mathcal K_N$.

For $b\in\R$, define $\mathbb G_N^b=\left\{g_{N,1}^b,\ldots,g_{N,r_N}^b\right\}$,
where $g_{N,j}^b(t)=g_{N,j}(t+b)$.
For all $\bx,\by\in X_N$, P-TFC gives
\[
 \bx\succsim_{\mathbb G_N^b}\by
 \Longleftrightarrow
 \bx+b\mathbf 1_N
 \succsim_N
 \by+b\mathbf 1_N
 \Longleftrightarrow
 \bx\succsim_N\by.
\]
Thus $\boldsymbol{\mathbb G}^b=\left(\mathbb G_N^b\right)_{N\in\mathcal N}$ and
$\boldsymbol{\mathbb G}$ are equivalent population-independent finite
representing systems. 
Then, Proposition 3 gives
\begin{equation}
 \operatorname{cone}\left\{\left[g_{N,1}^b\right],\ldots,\left[g_{N,r_N}^b\right]\right\}
 =
 \mathcal K_N.
 \label{eq:appendix-translated-cone}
\end{equation}
The map $\left[g\right]\mapsto\left[g(\,\cdot+b)\right]$ 
is invertible and linear on $C(\R)/\R\mathbf1$. 
By \eqref{eq:appendix-translated-cone}, it preserves
$\mathcal K_N$ and, in fact, leaves each of its extreme rays
invariant.\footnote{
    Denote the translation map $[g]\mapsto[g(\cdot+b)]$
    by $T_b$, and let $\sigma_b$ be its induced permutation
    of the $r_N$ extreme rays of $\mathcal K_N$.
    Since $T_b=\left(T_{b/\left(r_N\right)!}\right)^{\left(r_N\right)!}$, we have
    \[
      \sigma_b=\left(\sigma_{b/\left(r_N\right)!}\right)^{\left(r_N\right)!}=\operatorname{id},
    \]
    because the $r_N!$-th power of every permutation of 
    $r_N$ objects is the identity permutation.
}
Therefore, for every $b\in\R$ and $j=1,\ldots,r_N$,
there exist $a_{N,j}(b)>0$ and $c_{N,j}(b)\in\R$ such that
\[
 g_{N,j}(t+b)=a_{N,j}(b)g_{N,j}(t)+c_{N,j}(b)
 \qquad(t\in\R).
\]
This equation implies that the complete
ordering induced by each generator satisfies Translation-scale Full
Comparability. 
Therefore, Theorem 3 of Blackorby and Donaldson
\citeyearpar{Blackorby1982} implies that each $g_{N,j}$ is a
positive affine transformation of $\kappa_{\theta_{N,j}}$ for some
$\theta_{N,j}\geq0$.
Denoting $\Theta_N=\{\theta_{N,j}\mid j=1,\ldots,r_N\}$, we can conclude that
\[
 \langle\mathcal G_N\rangle
 =\operatorname{cone}\{\kappa_\theta\mid\theta\in\Theta_N\}
 +\operatorname{span}\{\mathbf1\}.
\]
This gives the representation in (ii).
Since the same generators are chosen for all populations,
$\Theta_{N}=\Theta_M$ for every $N,M\in\mathcal N$.
It remains to prove uniqueness. 
Suppose that $\boldsymbol{\Theta}'=\left(\Theta_N'\right)_{N\in\mathcal N}$ 
is another population-independent system of nonempty finite parameter sets 
giving the Kolm--Pollak representation of $\boldsymbol{\succsim}$. 
Applying Proposition 3, we have
\[ 
\operatorname{cone}\left\{\left[\kappa_\theta\right]\mid\theta\in\Theta_N\right\} = \operatorname{cone}\left\{\left[\kappa_\theta\right] \mid\theta\in\Theta_N'\right\} 
\quad\text{for every }N\in\mathcal N.
\]
For every finite set $A\subseteq\R_+$, the functions $\mathbf1$ and
$\left\{\kappa_\theta\right\}_{\theta\in A}$ are linearly independent in $C(\R)$. 
If there exists some $\theta\in\Theta_N\setminus\Theta_N'$, 
the above cone equality would express $\left[\kappa_\theta\right]$ as a nonnegative combination of
the classes indexed by $\Theta_N'$, 
contradicting this linear independence. 
Therefore, $\Theta_N=\Theta_N'$ for every $N$.
\qed

\medskip
\noindent
{\it Proof of Proposition 5.}
The implication (ii)$\Rightarrow$(i) is immediate,
since $\nu_\gamma(a\times\cdot)$ is a positive affine
transformation of $\nu_\gamma$ for every
$\gamma\leq1$ and $a>0$.

Suppose that (i) holds.
The analogue of Proposition 3 on $\R_{++}$ follows from the same proof.
Write $\left[g\right]=g+\R\mathbf1\in C\left(\R_{++}\right)/\R\mathbf1$.
As in the proof of Proposition 4, choose a
population-independent finite generating system
$\boldsymbol{\mathbb G}=\left(\mathbb G_N\right)_{N\in\mathcal N}$ representing
$\boldsymbol{\succsim}$, 
with $\mathbb G_N=\left\{g_{N,1},\ldots,g_{N,r_N}\right\}$,
whose classes represent the distinct extreme rays of
\[
 \mathcal K_N
 =\operatorname{cone}\left\{\left[g\right]\mid g\in\mathbb G_N\right\}.
\]

For $a>0$, consider a map $\left[g\right]\mapsto\left[g(a\,\cdot)\right]$. 
This map is invertible and linear on $C\left(\R_{++}\right)/\R\mathbf1$. 
By P-RFC and Proposition 3, this map preserves $\mathcal K_N$.
Since scaling by $a$ is the $r_N!$-fold composition of
scaling by $a^{1/(r_N!)}$, the permutation argument in
the proof of Proposition 4 shows that each extreme ray
is invariant.
Consequently, for every $j=1,\ldots,r_N$,
there exist $d_{N,j}(a)>0$ and $c_{N,j}(a)\in\R$
such that
\[
 g_{N,j}(at)
 =d_{N,j}(a)g_{N,j}(t)+c_{N,j}(a)
 \qquad(t>0).
\]
This equation implies that the complete ordering induced by each generator satisfies
Ratio-scale Full Comparability.  
Theorem 2 (i) of Blackorby and Donaldson
\citeyearpar{Blackorby1982} therefore implies that each
$g_{N,j}$ is a positive affine transformation of
$\nu_{\gamma_{N,j}}$ for some $\gamma_{N,j}\leq1$.
Setting
$\Gamma_N=\{\gamma_{N,j}\mid j=1,\ldots,r_N\}$
gives the representation in (ii).
Since the same generators are chosen for all populations, $\Gamma_{N}=\Gamma_M$ for all $N,M\in \mathcal{N}$.

Uniqueness follows as in the proof of Proposition 4,
using Proposition 3 on $\R_{++}$ and the linear
independence of $\mathbf1$ and
$\{\nu_\gamma\}_{\gamma\in A}$ for every finite set
$A\subseteq(-\infty,1]$.
\qed

\medskip
\noindent
{\it Proof of Proposition 6.}
Suppose that $\boldsymbol{\succsim}$ satisfies P-CFC.
Since P-CFC implies P-TFC, Proposition 4 yields a unique
population-independent system $\boldsymbol\Theta=(\Theta_N)_{N\in\mathcal N}$
of nonempty finite parameter sets giving the Kolm--Pollak representation.
For $a>0$, let $a\Theta_N=\{a\theta\mid\theta\in\Theta_N\}$.
Since
\[
 \kappa_\theta(at)=\kappa_{a\theta}(t)\quad(\theta>0),
 \qquad \kappa_0(at)=a\kappa_0(t),
\]
P-CFC implies that $(a\Theta_N)_{N\in\mathcal N}$ also represents
$\boldsymbol{\succsim}$.  This system is population independent, so
uniqueness gives $a\Theta_N=\Theta_N$ for every $a>0$.
Therefore, for each $N$, $\Theta_N=\{0\}$.

Conversely, the utilitarian criterion satisfies P-CFC because, 
for every $a>0$, $b\in\R$, and $\bx,\by\in X_N$,
\[
 \sum_{i\in N}(ax_i+b)-\sum_{i\in N}(ay_i+b)
 =a\left(\sum_{i\in N}x_i-\sum_{i\in N}y_i\right)
\]
holds.
\qed

\medskip
\noindent
\bibliography{literature} 

@book{AliprantisBorder2006,
  author    = {Aliprantis, Charalambos D. and Border, Kim C.},
  title     = {Infinite Dimensional Analysis: A Hitchhiker's Guide},
  publisher = {Springer},
  year      = {2006},
  edition   = {3},
  address   = {Berlin and Heidelberg},
  doi       = {10.1007/3-540-29587-9},
  isbn      = {978-3-540-29586-0}
}

@book{AdlerNorheim2022,
  editor    = {Matthew D. Adler and Ole F. Norheim},
  title     = {Prioritarianism in Practice},
  publisher = {Cambridge University Press},
  year      = {2022},
  doi       = {10.1017/9781108691734},
  isbn      = {9781108480932}
}

@book{Blackorby2005,
  author    = {Blackorby, Charles and Bossert, Walter and Donaldson, David J.},
  title     = {Population Issues in Social Choice Theory, Welfare Economics, and Ethics},
  publisher = {Cambridge University Press},
  year      = {2005},
  series    = {Econometric Society Monographs},
  address   = {Cambridge}
}

@article{BlackorbyDonaldson1977,
  author    = {Blackorby, Charles and Donaldson, David},
  title     = {Utility vs equity: Some plausible quasi-orderings},
  journal   = {Journal of Public Economics},
  year      = {1977},
  volume    = {7},
  number    = {3},
  pages     = {365--381},
  doi       = {10.1016/0047-2727(77)90055-X}
}

@article{Kamaga2018,
  author    = {Kamaga, Kohei},
  title     = {When do utilitarianism and egalitarianism agree on evaluation? An intersection approach},
  journal   = {Mathematical Social Sciences},
  year      = {2018},
  volume    = {94},
  pages     = {41--48},
  doi       = {10.1016/j.mathsocsci.2018.05.002}
}

@article{Blackorby1982,
  author    = {Blackorby, Charles and Donaldson, David},
  title     = {Ratio-Scale and Translation-Scale Full Interpersonal Comparability without Domain Restrictions: Admissible Social-Evaluation Functions},
  journal   = {International Economic Review},
  year      = {1982},
  volume    = {23},
  number    = {2},
  pages     = {249--268}
}

@techreport{Borie2016,
  author    = {Borie, Dino},
  title     = {Additively separable preferences without the completeness Axiom: An Algebraic Approach},
  institution = {GREDEG},
  year      = {2016},
  number    = {2016-11},
  type      = {Working Paper}
}

@article{Cato2018,
  author    = {Cato, Susumu},
  title     = {Incomplete decision-making and Arrow’s impossibility theorem},
  journal   = {Mathematical Social Sciences},
  year      = {2018},
  volume    = {94},
  pages     = {58--64},
  doi       = {10.1016/j.mathsocsci.2017.10.002}
}

@article{Cato2020,
author = {Cato, Susumu},
title = {Incompleteness, regularity, and collective preference},
journal = {Metroeconomica},
volume = {71},
number = {2},
pages = {333-344},
doi = {https://doi.org/10.1111/meca.12276},
url = {https://onlinelibrary.wiley.com/doi/abs/10.1111/meca.12276},
eprint = {https://onlinelibrary.wiley.com/doi/pdf/10.1111/meca.12276},
year = {2020}
}

@techreport{CerreiaVioglioOk2018,
  author      = {Cerreia-Vioglio, Simone and Ok, Efe A.},
  title       = {The Rational Core of Preference Relations},
  institution = {IGIER, Bocconi University},
  year        = {2018},
  number      = {632},
  type        = {Working Paper},
  url         = {https://igier.unibocconi.eu/publications/working-papers/rational-core-preference-relations}
}

@article{Danan2016,
  author    = {Danan, Eric and Gajdos, Thibault and Hill, Brian and Tallon, Jean-Marc},
  title     = {Robust social decisions},
  journal   = {American Economic Review},
  year      = {2016},
  volume    = {106},
  number    = {9},
  pages     = {2407--2425}
}

@article{Danan2015,
  author    = {Danan, Eric and Gajdos, Thibault and Tallon, Jean-Marc},
  title     = {Harsanyi's aggregation theorem with incomplete preferences},
  journal   = {American Economic Journal: Microeconomics},
  year      = {2015},
  volume    = {7},
  number    = {1},
  pages     = {61--69}
}

@techreport{Debreu1959,
  author    = {Debreu, Gerard},
  title     = {Topological Methods in Cardinal Utility Theory},
  institution = {Cowles Foundation for Research in Economics},
  year      = {1959},
  number    = {76},
  month     = {July}
}

@article{Dubra2004,
  author    = {Dubra, Juan and Maccheroni, Fabio and Ok, Efe A.},
  title     = {Expected utility theory without the completeness axiom},
  journal   = {Journal of Economic Theory},
  year      = {2004},
  volume    = {115},
  pages     = {118--133},
  doi       = {10.1016/S0022-0531(03)00166-2}
}

@article{Evren2011,
  author    = {Evren, {\"O}zg{\"u}r and Ok, Efe A.},
  title     = {On the multi-utility representation of preference relations},
  journal   = {Journal of Mathematical Economics},
  year      = {2011},
  volume    = {47},
  pages     = {554--563}
}

@article{Maskin1978,
  author    = {Maskin, Eric},
  title     = {A Theorem on Utilitarianism},
  journal   = {The Review of Economic Studies},
  year      = {1978},
  volume    = {45},
  number    = {1},
  pages     = {93--96}
}

@article{Nakamura2025,
  author  = {Nakamura, Kensei},
  title   = {When is it (im)possible to respect all individuals' preferences under uncertainty?},
  journal = {Games and Economic Behavior},
  year    = {2026},
  volume  = {158},
  pages   = {114--121},
  doi     = {10.1016/j.geb.2026.03.001}
}

@book{Sen2017,
  author    = {Sen, Amartya},
  title     = {Collective Choice and Social Welfare: An Expanded Edition},
  publisher = {Harvard University Press},
  year      = {2017},
  address   = {Cambridge, MA and London, England}
}

@article{Aumann1962,
  title={Utility theory without the completeness axiom},
  author={Aumann, Robert J},
  journal={Econometrica: Journal of the Econometric Society},
  pages={445--462},
  year={1962}
}

@article{Pivato2015,
  author  = {Marcus Pivato},
  title   = {Social choice with approximate interpersonal comparison of welfare gains},
  journal = {Theory and Decision},
  year    = {2015},
  volume  = {79},
  number  = {2},
  pages   = {181--216},
  doi     = {10.1007/s11238-014-9458-8}
}

@article{Pivato2013,
  author  = {Marcus Pivato},
  title   = {Social welfare with incomplete ordinal interpersonal comparisons},
  journal = {Journal of Mathematical Economics},
  year    = {2013},
  volume  = {49},
  number  = {5},
  pages   = {405--417},
  doi     = {10.1016/j.jmateco.2013.04.004}
}

@misc{ChungMiyagishima2024,
  author       = {Hun Chung and Kaname Miyagishima},
  title        = {The Priority Axiom: A New Foundation for Prioritarianism},
  year         = {2024},
  howpublished = {Working paper},
  doi          = {10.13140/RG.2.2.14049.28008},
}

@misc{Parfit1991,
  author       = {Derek Parfit},
  title        = {Equality or Priority?},
  year         = {1991},
  month        = nov,
  day          = {21},
  howpublished = {The Lindley Lecture, Department of Philosophy, University of Kansas},
  institution  = {University of Kansas, Department of Philosophy},
  note         = {Lindley Lecture manuscript}
}

@incollection{Alesina2011,
title = {Chapter 4 - Preferences for Redistribution},
editor = {Jess Benhabib and Alberto Bisin and Matthew O. Jackson},
series = {Handbook of Social Economics},
publisher = {North-Holland},
volume = {1},
pages = {93-131},
year = {2011},
issn = {1570-6435},
doi = {https://doi.org/10.1016/B978-0-444-53187-2.00004-8},
url = {https://www.sciencedirect.com/science/article/pii/B9780444531872000048},
author = {Alberto Alesina and Paola Giuliano}
}

@book{Rockafellar1970,
  author    = {R. Tyrrell Rockafellar},
  title     = {Convex Analysis},
  series    = {Princeton Mathematical Series},
  volume    = {28},
  publisher = {Princeton University Press},
  address   = {Princeton, NJ},
  year      = {1970}
}

@article{McCarthy2021,
  author  = {David McCarthy and Kalle Mikkola and Teruji Thomas},
  title   = {Expected Utility Theory on Mixture Spaces without the Completeness Axiom},
  journal = {Journal of Mathematical Economics},
  volume  = {97},
  pages   = {102538},
  year    = {2021}
}

@article{Muliere1989,
  author  = {Muliere, Pietro and Scarsini, Marco},
  title   = {A note on stochastic dominance and inequality measures},
  journal = {Journal of Economic Theory},
  volume  = {49},
  pages   = {314--323},
  year    = {1989},
  doi     = {10.1016/0022-0531(89)90084-7}
}

@article{Thistle1989,
  author  = {Thistle, Paul D.},
  title   = {Ranking distributions with generalized {L}orenz curves},
  journal = {Southern Economic Journal},
  volume  = {56},
  number  = {1},
  pages   = {1--12},
  year    = {1989}
}
\bibliographystyle{apalike}

\end{document}